\documentclass[11pt,a4paper]{article}
\pdfoutput=1

\usepackage[colorlinks=true, linkcolor=black!50!blue, urlcolor=blue, citecolor=blue, anchorcolor=blue]{hyperref}
\usepackage[font=small,labelfont=bf,margin=0mm,labelsep=period,tableposition=top]{caption}
\usepackage[a4paper,top=1.5cm,bottom=2cm,left=1.5cm,right=1.5cm,bindingoffset=0mm]{geometry}

\usepackage{placeins,cite}
\usepackage{graphicx}
\usepackage{soul}
\usepackage{subcaption}
\usepackage{float}
\usepackage{afterpage}
\usepackage{epsfig}
\usepackage{amssymb}
\usepackage{amsmath}
\usepackage{bm}
\usepackage{multirow}
\usepackage{url}
\usepackage{xcolor}
\usepackage{ulem}
\usepackage{url}
\usepackage{booktabs,multirow}
\usepackage{textcomp}
\usepackage{comment}
\graphicspath{{../}{./figures/}}

\usepackage{makecell}

\makeatletter
\def\thickhline{%
             \noalign{\ifnum0 =`}\fi\hrule \@height \thickarrayrulewidth \futurelet
             \reserved@a\@xthickhline}
\def\@xthickhline{\ifx\reserved@a\thickhline
                \vskip\doublerulesep
                \vskip -\thickarrayrulewidth
                \fi
                \ifnum0 =`{\fi}}
\makeatother
\newlength{\thickarrayrulewidth}
\usepackage{tikz}
\usetikzlibrary{positioning}
\usetikzlibrary{shapes.geometric, arrows}
\usetikzlibrary{arrows.meta}
\usepackage{varwidth}
\usepackage{xcolor}
\definecolor{mtplotlib1}{HTML}{1f77b4}
\definecolor{mtplotlib2}{HTML}{ff7f0e}
\definecolor{mtplotlib3}{HTML}{2ca02c}
\definecolor{mtplotlib4}{HTML}{d62728}

\tikzset{%
  >={Latex[width=2mm,length=2mm]},
            base/.style = {rectangle, rounded corners, draw=black,
                           minimum width=4cm, minimum height=1cm,
                           text centered}, 
            mystyle/.style={rectangle, rounded corners, draw=black,
            minimum width=12cm, minimum height=1cm,
            text centered}, 
    col0/.style = {base, fill=white!30},
    col1/.style = {base, fill=mtplotlib1!30},
    col11/.style = {mystyle, fill=mtplotlib1!30},
    col2/.style = {base, fill=mtplotlib2!30},
    col3/.style = {base, fill=mtplotlib3!30},
    col4/.style = {base, minimum width=2.5cm, fill=mtplotlib4!15,}
}

\newcommand{\POWHEG}{{\sc\small POWHEG}}
\newcommand{\be}{\begin{equation}}
\newcommand{\ee}{\end{equation}}
\newcommand{\bea}{\begin{eqnarray}}
\newcommand{\eea}{\end{eqnarray}}
\newcommand{\bi}{\begin{itemize}}
\newcommand{\ei}{\end{itemize}}
\newcommand{\ben}{\begin{enumerate}}
\newcommand{\een}{\end{enumerate}}

\newcommand{\lc}{\left[}
\newcommand{\rc}{\right]}
\newcommand{\lp}{\left(}
\newcommand{\rp}{\right)}

\def\frac#1#2{{{#1}\over {#2}}}
\def\gsim{\mathrel{\rlap{\lower4pt\hbox{\hskip1pt$\sim$}}
    \raise1pt\hbox{$>$}}}       
\def\lsim{\mathrel{\rlap{\lower4pt\hbox{\hskip1pt$\sim$}}
    \raise1pt\hbox{$<$}}}

\newcommand{\draft}[1]{}
\def\beq{\begin{equation}}
\def\eeq{\end{equation}}

\numberwithin{equation}{section}
\numberwithin{figure}{section}
\numberwithin{table}{section}

\usepackage{tabularx}
\newcolumntype{C}[1]{>{\centering\arraybackslash}p{#1}}

\definecolor{darkblue}{rgb}{0.0,0,0.5}
\definecolor{darkgreen}{rgb}{0.0,0.3,0.0}
\definecolor{redish}{rgb}{0.675,0,0.2}
\definecolor{red}{rgb}{0.8,0,0}
\definecolor{green}{rgb}{0,0.6,0}
\definecolor{bluish}{rgb}{0.2,0.2,0.675}
\definecolor{mygrey}{rgb}{0.6,0.6,0.6}

\usepackage{tikz} 
\usetikzlibrary{shapes,arrows,positioning,automata,backgrounds,calc,er,patterns,arrows.meta}
\usepackage{tikz-feynman}
\tikzfeynmanset{compat=1.0.0}
\usepackage{varwidth}
\usepackage{xcolor}
\definecolor{mtplotlib1}{HTML}{1f77b4}
\definecolor{mtplotlib2}{HTML}{ff7f0e}
\definecolor{mtplotlib3}{HTML}{2ca02c}
\definecolor{mtplotlib4}{HTML}{d62728}

\tikzset{%
  >={Latex[width=2mm,length=2mm]},
            base/.style = {rectangle, rounded corners, draw=black,
                           minimum width=4cm, minimum height=1cm,
                           text centered}, 
            mystyle/.style={rectangle, rounded corners, draw=black,
            minimum width=12cm, minimum height=1cm,
            text centered}, 
    col0/.style = {base, fill=white!30},
    col1/.style = {base, fill=mtplotlib1!30},
    col11/.style = {mystyle, fill=mtplotlib1!30},
    col2/.style = {base, fill=mtplotlib2!30},
    col3/.style = {base, fill=mtplotlib3!30},
    col4/.style = {base, minimum width=2.5cm, fill=mtplotlib4!15,}
}

\usepackage{tabularx}
\usepackage{subcaption}
\newcolumntype{C}[1]{>{\centering\arraybackslash}p{#1}}

\definecolor{lightblue}{rgb}{0.0,0.5,1.0}

\begin{document}
\newgeometry{top=1.5cm,bottom=1.5cm,left=1.5cm,right=1.5cm,bindingoffset=0mm}

\vspace{-2.0cm}
\begin{flushright}
   Nikhef 2024-012\\
CERN-TH-2024-111
\end{flushright}
\vspace{0.6cm}

\begin{center}
  {\Large \bf A Phenomenological Analysis of LHC Neutrino Scattering  \\[0.3cm] at NLO Accuracy Matched to Parton Showers }\\
  \vspace{1.1cm}
  {\small
  Melissa van Beekveld$^{1}$, Silvia Ferrario Ravasio$^{2}$, Eva Groenendijk$^{1}$, Peter Krack$^{1,3}$, \\[0.1cm] Juan Rojo$^{1,3}$, and Valentina Sch\"utze Sánchez$^{1}$
  }\\
  
\vspace{0.7cm}

{\it \small
  ~$^1$Nikhef Theory Group, Science Park 105, 1098 XG Amsterdam, The Netherlands\\[0.1cm]
  ~$^2$Theoretical Physics Department, CERN, Esplanade des Particules 1, Meyrin, Switzerland\\[0.1cm]  
    ~$^3$Department of Physics and Astronomy, Vrije Universiteit, NL-1081 HV Amsterdam\\[0.1cm]
   
 }

\vspace{1.0cm}

{\bf \large Abstract}

\end{center}

We perform a detailed phenomenological study of high-energy
neutrino deep inelastic scattering (DIS) focused on LHC far-forward experiments such as FASER$\nu$ and SND@LHC.
To this aim, we parametrise the neutrino fluxes reaching these LHC far-forward experiments in terms of `neutrino
PDFs' encoding their energy and rapidity dependence by means of the
{\sc\small LHAPDF} framework.
We integrate these neutrino PDFs in the recently developed {\sc\small POWHEG-BOX-RES}  implementation of neutrino-induced DIS to produce predictions accurate at next-to-leading order (NLO) in the QCD coupling matched to parton showers (PS) with {\sc\small Pythia8}.
We present NLO+PS predictions for final-state distributions within the acceptance for FASER$\nu$ and SND@LHC as well as for two experiments of the proposed Forward Physics Facility (FPF), FASER$\nu$2 and FLArE.
We quantify the impact of NLO QCD corrections, of the parton showering and hadronisation settings in {\sc\small Pythia8}, of the QED shower, and of the incoming neutrino flavour for the description of these observables, and compare our predictions with the {\sc\small GENIE} neutrino event generator.
Our work demonstrates the relevance of modern higher-order event generators to achieve the key scientific targets of the LHC neutrino experiments.
  
\clearpage

\tableofcontents

\section{Introduction}

The recent groundbreaking observation of LHC neutrinos from the FASER~\cite{FASER:2023zcr} and SND@LHC~\cite{SNDLHC:2023pun} far-forward experiments, together with the first measurement of the neutrino interaction cross-section at TeV energies by FASER$\nu$~\cite{FASER:2024hoe}, herald the beginning of the collider neutrino era in particle physics.
LHC neutrinos, characterised by the highest energies ever achieved in a laboratory experiment, provide unique information enabling novel opportunities in various areas of particle, hadronic, and astroparticle physics (see~\cite{Anchordoqui:2021ghd,Feng:2022inv} for an overview).

Both FASER/FASER$\nu$ and SND@LHC will take data for the rest of Run-3, and FASER has also been approved to operate (albeit without the FASER$\nu$  module) during Run-4.
An upgrade of FASER$\nu$, dubbed FASER$\nu$2,
together with new far-forward experiments, such as FLArE~\cite{Batell:2021blf}, may be installed in a proposed Forward Physics Facility (FPF)~\cite{Anchordoqui:2021ghd,Feng:2022inv} operating concurrently with the HL-LHC.
In addition, the upgrade of SND@LHC, known as AdvSND, has been proposed to operate in a different location in the CERN complex during Run-4 and potentially beyond.

A crucial ingredient for the robust interpretation of LHC far-forward neutrino experiments is the accurate simulation of the underlying neutrino-hadron interactions, namely the deep-inelastic scattering (DIS) process, including the initial- and final-state radiation associated to the leptonic and hadronic final states.
While inclusive cross-sections and double-differential distributions in $x$ (partonic momentum fraction) and $Q^2$ (momentum transfer between the neutrino and the target) can be reliably evaluated using fixed-order QCD calculations up to $\mathcal{O}(\alpha_s^3)$ (N$^3$LO)~
\cite{Moch:2004xu,Vermaseren:2005qc,Moch:2008fj,Blumlein:2022gpp,Currie:2018fgr, Gehrmann:2018odt,Karlberg:2024hnl,McGowan:2022nag,NNPDF:2024nan}
in the perturbative expansion, more exclusive distributions accounting for realistic acceptance cuts or sensitive to the details of hadronic final states need to be modelled by means of parton shower (PS) Monte Carlo (MC) event generators~\cite{Campbell:2022qmc}.
While progress in event generation for proton-proton collisions at the LHC has been spectacular in recent years, somewhat less attention has been devoted to the corresponding description of lepton-nucleon collisions.
The situation is now rapidly changing, partly motivated by the upcoming availability of the Electron-Ion Collider (EIC)~\cite{AbdulKhalek:2021gbh} to start taking data in the early 2030s.

In the case of neutrino scattering experiments, both at colliders and in the context of atmospheric and astroparticle neutrino physics, many analyses rely on leading order (LO) event generators such as {\sc\small GENIE}~\cite{Andreopoulos:2009rq}, where LO matrix elements with effective PDFs from the Bodek-Yang model~\cite{Yang:1998zb} are matched to {\sc\small Pythia6}~\cite{Sjostrand:2000wi} for the parton shower and hadronisation.
Restricted to inclusive fixed-order calculations, {\sc\small GENIE} can be extended to NLO and NNLO structure functions with user-defined parton distribution functions (PDFs)~\cite{Gao:2017yyd} via the {\sc\small HEDIS} module~\cite{Garcia:2020jwr}, while extension to the low-$Q$ regime (shallow inelastic scattering or SIS) is possible by means of the data-driven NNSF$\nu$ framework~\cite{Candido:2023utz}.
The higher-order QCD calculations implemented in {\sc\small GENIE}, however, are applicable only to inclusive cross-sections and double-differential distributions without acceptance cuts, and they do not provide the exclusive modelling of the final state of the interaction since they are not matched to a PS generator. 
Beyond LO, the general-purpose Monte Carlo event generators {\sc Sherpa}~\cite{Sherpa:2019gpd} and {\sc Herwig7}~\cite{Bellm:2015jjp,Bewick:2023tfi} allow for DIS simulations at NLO+PS, and also recent DIS event generators~\cite{Banfi:2023mhz,Borsa:2024rmh,Buonocore:2024pdv,FerrarioRavasio:2024kem} employing the \POWHEG~method \cite{Nason:2004rx,Frixione:2007vw,Alioli:2010xd} enable NLO+PS predictions.
NNLO+PS predictions for charged-current DIS are also available in a private version of the {\sc Sherpa} framework~\cite{Hoche:2018gti}.

The goal of this work is to perform a detailed phenomenological study of neutrino-induced DIS processes as measured at the LHC far-forward experiments, accounting for NLO QCD corrections and modern parton shower and hadronisation algorithms.
To this aim, we construct dedicated `neutrino PDFs' describing the forward flux of incoming neutrinos.
The energy and rapidity dependence of these neutrino fluxes produced at the ATLAS interaction point (IP) are taken from recent calculations~\cite{Kling:2021gos,FASER:2024ykc,Buonocore:2023kna} and parameterised by means of the {\sc\small LHAPDF} interface~\cite{Buckley:2014ana}.
We integrate the neutrino PDFs with the {\sc\small POWHEG-BOX-RES}~\cite{Jezo:2015aia} event generator presented in~\cite{Banfi:2023mhz,FerrarioRavasio:2024kem}, enabling the combination of NLO-accurate DIS calculations with parton showering, hadronisation, and other non-perturbative QCD phenomena from modern MC generators such as {\sc\small Pythia8}~\cite{Sjostrand:2014zea}.
Being made available via {\sc\small LHAPDF},
the neutrino PDFs used in our  framework may be easily used for other event generators modelling LHC neutrino scattering.

We first validate the \POWHEG~neutrino DIS simulations at LO and NLO in the fixed energy case with 
{\sc\small Pythia8} LO simulations and with inclusive fixed-order NLO calculations provided by {\sc\small YADISM}~\cite{Candido:2024rkr}, respectively, finding good agreement. 
We also compare the \POWHEG~predictions with those from {\sc\small GENIE}.
We then present differential NLO+PS predictions 
for final-state distributions, including acceptance cuts, of phenomenological relevance for the interpretation of the FASER$\nu$
and SND@LHC measurements, and consider also the FPF experiments FASER$\nu$2 and FLArE.
We verify that the event yields expected for the LHC neutrino experiments derived in~\cite{Cruz-Martinez:2023sdv} are reproduced by the \POWHEG-based simulations combined with the neutrino PDFs.
We quantify the impact of NLO QCD corrections, showing that these are in general sizeable for differential distributions, and 
assess the impact of the  parton shower model by considering both the {\sc\small Pythia8} shower with fully local dipole recoil~\cite{Cabouat:2017rzi} and {\sc\small Vincia}~\cite{Fischer:2016vfv}, as well as of the choice of the soft QCD physics tune. 

While our draft was being finalised,  another \POWHEG-based generator for DIS, including heavy quark mass effects, was presented~\cite{Buonocore:2024pdv}.
For the phenomenological studies carried out here, such quark mass effects are small, and the generators in~\cite{FerrarioRavasio:2024kem,Buonocore:2024pdv} should be close for inclusive scattering.
We also point out that our treatment of the incoming lepton flux through neutrino PDFs could in principle be applied also to the generator presented in~\cite{Buonocore:2024pdv}.  

The outline of this paper is as follows. 
First, in Sect.~\ref{sec:neutrinoPDF} we lay out the notation, define the observables in far-forward neutrino experiments for which theoretical predictions will be obtained using the  \POWHEG~generators introduced in~\cite{Banfi:2023mhz,FerrarioRavasio:2024kem}, and construct the neutrino PDFs parameterising the energy and flavour dependence of the LHC neutrino fluxes.
We benchmark the output of \POWHEG~in the fixed neutrino energy case with the {\sc\small YADISM} and stand-alone {\sc\small Pythia8} predictions in Sect.~\ref{sec:results_validation}, and also compare these results with those from {\sc\small GENIE}.
Predictions for differential final-state distributions within the FASER$\nu$, SND@LHC, FASER$\nu$2, and FLArE acceptances, accurate at  NLO+PS, are presented in Sect~\ref{sec:results}.
Finally, in Sect.~\ref{sec:summary} we summarise the main findings of our work and discuss the outlook for possible future applications.

The {\sc\small POWHEG-BOX-RES} neutrino DIS code presented in~\cite{Banfi:2023mhz,Borsa:2024rmh,FerrarioRavasio:2024kem} and used to obtain the results of this work can be obtained from {\tt svn://powhegbox.mib.infn.it/trunk/User-Processes-RES/DIS}.
Neutrino PDF grid files and example run cards to reproduce the results of this paper are also provided there.

\section{Neutrino PDFs for event generation}
\label{sec:neutrinoPDF}

Here first we review the main ingredients of the neutrino-initiated DIS processes and define the physical observables for which we will provide theoretical predictions.
We then describe how to encapsulate the LHC neutrino fluxes~\cite{Kling:2021gos,Buonocore:2023kna,FASER:2024ykc} into a formalism equivalent to neutrino PDFs in analogy with the quark and gluon PDFs in the proton.
We validate our implementation of neutrino PDFs in {\sc\small POWHEG} by reproducing the independent calculation of neutrino event yields at the LHC far-forward detectors of~\cite{Cruz-Martinez:2023sdv}.

\subsection{Neutrino deep-inelastic scattering}
\label{subsec:nudis}

\paragraph{Kinematics.}
We are interested in neutrino-initiated deep-inelastic scattering,
\bea
\label{eq:process_def_1}
\nu_\ell(p_\nu) + N(p_N) &\to& \ell^-(p_\ell) + X_h(p_h) \,, \qquad \ell=e, \mu, \tau\, \quad ({\rm charged\,current)} \, , \\
\nu_\ell(p_\nu) + N(p_N) &\to& \nu_\ell(p'_\nu)  + X_h(p_h) \,, \qquad \ell=e, \mu, \tau\, \quad {\rm (neutral\,current)} \, ,
\eea
and on the corresponding reactions for processes initiated by antineutrinos,
\bea
\bar{\nu}_\ell(p_\nu) + N(p_N) &\to& \ell^+(p_\ell) + X_h(p_h) \,, \qquad \ell=e, \mu, \tau\, \quad ({\rm charged\,current)} \, , \\
\bar{\nu}_\ell(p_\nu) + N(p_N) &\to& \bar{\nu}_\ell(p'_\nu)  + X_h(p_h) \,, \qquad \ell=e, \mu, \tau\, \quad {\rm (neutral\,current)} 
\,  ,
\label{eq:process_def_2}
\eea
where $N$ indicates a generic nucleon, either proton ($N
=p$) or the average nucleon in a heavy nuclear target such as argon (${\rm Ar}$) or tungsten (${\rm W}$).
The hadronic final state is indicated by $X_h$ and the four-momenta of each particle species $X$ by $p_X$.
The charged-current neutrino (antineutrino) process is mediated by a $W^-$ ($W^+$) boson exchange, while the neutral-current scattering is mediated by a $Z^0$-boson exchange.

In this work we consider both the laboratory frame (also known as fixed-target frame), with the nucleons in the target at rest,
\be
\label{eq:labframe_kinematics}
p^{\mu}_{\nu} = (E_\nu, 0, 0, E_\nu)\, ,
\qquad p^{\mu}_N = (m_N, 0, 0,0) \, ,
\ee
with $m_N$ being the nucleon mass, as well as the collider frame defined by
\be
p^{\mu}_{\nu} = (\widetilde{E}_\nu, 0, 0, \widetilde{E}_\nu)\, ,
\qquad p^{\mu}_N = (E_N, 0, 0,-p_{N,z}) \, ,
\ee
where $p_{N,z}= E_N$ neglecting nucleon mass effects.
While results for Lorentz-invariant quantities are frame-independent, experimentally one is interested in frame-specific quantities such as the energy or the angle of the outgoing lepton in the fixed-target frame. 
Here we will present results for {\sc\small POWHEG} simulations in the laboratory frame, and we have verified that equivalent results are obtained in the collider frame for  frame-independent variables such as the momentum transfer squared $Q^2$ and the partonic momentum fraction $x$ for the same value of the centre-of-mass energy.

Working therefore in the laboratory frame and focusing on the charged-current process, we will be mostly interested in the energies of the charged lepton, $E_\ell$, its scattering angle  $\theta_\ell$ with respect to the incoming neutrino axis (which defines the $z$-axis), and the total energy of the hadronic system $E_h$,
\be
\label{eq:final_state_kinematics}
p^{\mu}_{\ell} = (E_\ell, E_\ell \sin\theta_\ell \cos\phi_\ell, E_\ell \sin\theta_\ell \sin\phi_\ell, E_\ell \cos\theta_\ell) \, ,
\qquad p^{\mu}_h = (E_h, \vec{p}_h) \, ,
\ee
where $\phi_\ell$ is the azimuthal angle of the charged lepton.
The frame-independent DIS variables $Q^2$, $x$, and $y$ (inelasticity) are constructed from the various four-momenta as 
\be
Q^2 = -q^2 = -(p_\nu^\mu - p_\ell^\mu)^2\, ,\quad  
x = \frac{Q^2}{2p_N \cdot q} \, ,\quad 
y = \frac{p_N \cdot q}{p_N \cdot p_\nu} \, .
\ee
The measurement of the three independent kinematic variables (only two in the case of LO calculations) enables the reconstruction of the DIS kinematics.
For instance, in charged-current neutrino DIS, a measurement of $\lp E_\ell,\theta_\ell, E_h \rp$ fixes the DIS kinematics to be
 \bea
 E_\nu &=& E_h + E_\ell \, , \nonumber \\
 Q^2 &=& 4 ( E_h + E_\ell) E_\ell \sin^2 \lp \theta_\ell/2\rp \, ,  \label{eq:dis_kinematic_mapping}\\
 x&=& \frac{4 ( E_h + E_\ell) E_\ell \sin^2 \lp \theta_\ell/2\rp}{2m_N E_h} \, .\nonumber
 \eea
 The value of the inelasticity $y$ then follows from the obtained values of $E_\nu$, $Q^2$, and $x$,
 \be
y = \frac{Q^2}{x s}= \frac{Q^2}{2xm_N E_\nu} \, ,
 \ee
 with $s=\lp p_\nu + p_N\rp^2$ and where the last equality applies only in the laboratory frame. 
 In this frame, one sees that the inelasticity
 \be
y = \frac{E_h}{E_h+E_\ell} \le 1 \, ,
 \ee
 quantifies the fraction of the initial neutrino energy that is transferred to the hadronic final state.
Finally, the invariant mass of the hadronic final state $W^2$ is given by
\be
W^2 = \lp p^\mu_h\rp^2 = m_N^2 + Q^2\frac{(1-x)}{x} \, .
\ee
A cut in $W^2$ ($\gsim 4~{\rm GeV}^2$) is required to ensure that neutrino scattering takes place in the deep-inelastic, perturbative region. 

\paragraph{Structure functions.}
In the following we assume the target to be composed of protons, $N=p$.
 Extension to other targets only modifies the underlying partonic decomposition.
 The differential cross-section for charged-current scattering
can be expressed in terms of three
independent structure functions~\cite{10.1093/ptep/ptac097}
\be
\label{eq:neutrino_DIS_xsec_FL_Q2}
\frac{d^2\sigma^{\nu p}(x,Q^2,y)}{dxdQ^2} =  \frac{G_F^2}{4\pi x\lp 1+Q^2/m_W^2\rp^2}\lc Y_+F^{\nu p}_2(x,Q^2) - y^2F^{\nu p}_L(x,Q^2) +Y_- xF^{\nu p}_3(x,Q^2)\rc  \, ,
\ee
\be
\label{eq:antineutrino_DIS_xsec_FL_Q2}
\frac{d^2\sigma^{\bar{\nu} p}(x,Q^2,y)}{dxdQ^2} =  \frac{G_F^2}{4\pi x\lp 1+Q^2/m_W^2\rp^2}\lc Y_+F^{\bar{\nu} p}_2(x,Q^2) - y^2F^{\bar{\nu} p}_L(x,Q^2) -Y_- xF^{\bar{\nu} p}_3(x,Q^2)\rc  \, ,
\ee
for neutrino and antineutrino scattering, respectively and with $Y_\pm = 1\pm \lp 1-y\rp^2$. 
Eqns.~(\ref{eq:neutrino_DIS_xsec_FL_Q2}) and~(\ref{eq:antineutrino_DIS_xsec_FL_Q2}) are valid provided
the hadronic 
invariant mass $W$  is above the resonance production threshold, $W \gsim 2$ GeV.
In the perturbative regime, the neutrino DIS structure functions
are expressed as
\be
\label{eq:sfs_pqcd}
 F^{\nu p}_i(x,Q^2) = \sum_{j=q,\bar{q},g}\int_x^1 \frac{dz}{z}\, C_{i,j}^{\nu N}(z,\alpha_s(Q^2))f_j^{(p)}\lp \frac{x}{z},Q^2\rp \, , \quad i = 2,3,L \, ,
 \ee
i.e.\ as a convolution of partonic coefficient functions  $C_{i,j}^{\nu N}(x,\alpha_s)$ and
process-independent proton PDFs $f^{(p)}_j\lp x,Q^2\rp$.
Experimental results are often presented in terms of dimensionless reduced cross-sections~\cite{H1:2012qti} where numerical prefactors cancel out and the dependence on the DIS structure functions is exposed, 
\begin{equation}
    \sigma_{\rm R}^{\nu p}(x,Q^2,y) = \frac{4\pi x (m_W^2 + Q^2)^2}{G_F^2 m_W^4} \frac{d^2 \sigma^{\nu p}}{dx dQ^2} =\lc Y_+F^{\nu p}_2(x,Q^2) - y^2F^{\nu p}_L(x,Q^2) +Y_- xF^{\nu p}_3(x,Q^2)\rc \, .
    \label{eq:sigmaR_CC}
\end{equation}
Note that this definition of the charged-current reduced cross-section has a factor 2 difference as compared to the  case of charged-lepton initiated scattering.

\subsection{Neutrino PDFs for DIS event generation}
\label{subsec:general_formalism}

In the context of a Monte Carlo event generator tailored to the LHC far-forward experiments, accounting for the neutrino fluxes in terms of neutrino PDFs has two main advantages.
First, one can deploy the {\sc\small LHAPDF} framework~\cite{Buckley:2014ana} to parameterise and interpolate the energy dependence of the predictions for the neutrino fluxes as well as their associated uncertainties (variations).
Second, in this approach, neutrino-proton scattering shares many similarities with the structure of proton-proton or proton-ion collisions, hence leveraging existing functionalities available in LHC event generators such as automated PDF reweighting.

We assume a LHC far-forward detector of length $L_T$ exposed to a flux of neutrinos generated at the ATLAS interaction point given by $dN_{\nu_i}/dE_\nu$, with $i=e,\mu,\tau$ indicating the neutrino flavour (and likewise for antineutrinos, $dN_{\bar{\nu}_i}/dE_\nu$) and $E_{\nu}$ its energy in the laboratory frame, Eq.~(\ref{eq:labframe_kinematics}).
The magnitude of this flux depends on the neutrino flavour, the cross-sectional geometry of the detector, and its position with respect to the line of sight (LoS) of the proton beams. 
The calculation of $dN_{\nu_i}/dE_\nu$ also assumes a given integrated luminosity $\mathcal{L}_{pp}$ from the primary proton-proton collisions at the IP.

As discussed in Sect.~\ref{subsec:nudis}, the measurement of the final-state kinematic variables $E_\ell$, $\theta_\ell$, and $E_h$, see also Eq.~(\ref{eq:final_state_kinematics}), in charged-current neutrino-hadron scattering allows reconstructing the DIS kinematic variables $x$ and $Q^2$ as well as the neutrino energy $E_\nu$.
Following~\cite{Cruz-Martinez:2023sdv}, one can express $N_{\rm int}^{(\nu_i)}$, the number of neutrinos of $i-$th flavour interacting via charged-current DIS, for a given detector geometry, incoming neutrino flux, and kinematic range  as follows
\be
  \label{eq:event_yields_calculation}
   N_{\rm int}^{(\nu_i)}= n_T L_T\int_{Q^{2}_{\rm min}}^{Q^{2}_{\rm max}}\int_{x_{\rm min}}^{x_{\rm max}}\int_{E_{\rm min}}^{E_{\rm max}} dQ^2 dx dE_{\nu}\,\frac{dN_{\nu_i}(E_\nu)}{dE_{\nu}} \left(\frac{d^2\sigma^{\nu_i A}(x,Q^2,E_{\nu})}{dxdQ^2}\right) {\cal A}(E_\ell,\theta_\ell,E_h)  \, , 
\ee
with $n_T$ being the nucleon density of the target detector material, $d^2\sigma^{\nu_i A}/dxdQ^2$ the DIS double-differential cross-section Eq.~(\ref{eq:neutrino_DIS_xsec_FL_Q2}) for a nuclear target $A$, and we integrate from the minimum to maximum values of the DIS variables $x$, $Q^2$ and of the neutrino energy $E_\nu$ considered in the binned measurement. 

In Eq.~(\ref{eq:event_yields_calculation}),  ${\cal A}(E_\ell,\theta_\ell,E_h)$ indicates an acceptance factor which takes the form of multiple step functions and accounts for the experimental acceptances in the final-state kinematic variables $E_\ell,\ \theta_\ell$ and $E_h$.
Within the context of an event generator such as {\sc\small POWHEG}, accounting for acceptance corrections at the generation level in the form of Eq.~(\ref{eq:event_yields_calculation}) is only possible at LO, since in this case one can combine this acceptance factor directly with the LO matrix element before the MC integration takes place.
For event generation at N$^k$LO accuracy with $k\ge 1$, this approach is not feasible and experimental acceptances can be introduced instead at the event analysis level.
Eq.~(\ref{eq:event_yields_calculation}) holds to all orders in the QCD expansion if the DIS cross-section is evaluated within the structure function formalism.

For neutrinos produced in symmetric proton-proton collisions at the LHC with a centre-of-mass energy of $\sqrt{s_{\rm pp}}=2E_p$, the neutrino energy must satisfy $E_\nu \le E_p$. 
In analogy with the quark and gluon PDFs, we define a ``neutrino momentum fraction'' $x_{\nu}$ as
\be
x_{\nu} \equiv \frac{E_\nu}{\sqrt{s_{\rm pp}}} \, ,\qquad 0 \le x_{\nu} \le 1 \, ,
\label{eq:xnu}
\ee
and consequently encapsulate the neutrino flux in terms of a scale-independent ``neutrino PDF'' defined as
\be
\label{eq:neutrino_pdf_definition}
f_{\nu_i}(x_{\nu})\equiv \sqrt{s_{\rm pp}} \frac{dN_{\nu_i}(E_\nu)}{dE_{\nu}} \, ,\qquad i = e,\mu,\tau \, .
\ee
The normalisation of Eq.~(\ref{eq:neutrino_pdf_definition}) has been chosen such that the neutrino PDFs are dimensionless, and that the ``sum rule'' associated to this neutrino PDF is given by
\be
\label{eq:neutrinoPDF_sumrule}
\int_0^1 dx_{\nu} f_{\nu_i}(x_{\nu})
= \int_0^{\sqrt{s_{\rm pp}}} dE_{\nu}
\frac{dN_{\nu_i}(E_\nu)}{dE_{\nu}} =  N_{\rm flux}^{(\nu_i)} \, ,\qquad i = e,\mu,\tau \, ,
\ee
with $N_{\rm flux}^{(\nu_i)}$
denoting the total number of neutrinos of a given flavour expected to travel across the cross-sectional area of the considered detector for a given integrated luminosity $\mathcal{L}_{\rm pp}$ in the primary proton-proton collision.
Here for simplicity we take the normalisation factor in Eq.~(\ref{eq:neutrino_pdf_definition}) to be $\sqrt{s_{\rm pp}}=14$ TeV for all far-forward detectors considered, though for the Run-3 experiments the neutrino fluxes that we use have been evaluated with 13.6 TeV as centre-of-mass energy.
Note that $N_{\rm flux}^{(\nu_i)}$ should not be confused with $N_{\rm int}^{(\nu_i)}$ from Eq.~(\ref{eq:event_yields_calculation}), which instead corresponds to the total number of neutrinos interacting within acceptance of a specific detector. 
   
Expressing the event yield calculation Eq.~(\ref{eq:event_yields_calculation}) in terms of the neutrino PDFs leads to
\be
\label{eq:event_yields_calculation_nuPDF}
   N_{\rm int}^{(\nu_i)}= n_T L_T\int_{Q^{2}_{\rm min}}^{Q^{2}_{\rm max}}dQ^2 \int_{x_{\rm min}}^{x_{\rm max}} dx \int_{0}^{1}  dx_{\nu}   f_{\nu_i}(x_{\nu}) \left(\frac{d^2\sigma^{\nu_i A}(x,Q^2,x_\nu\sqrt{s_{\rm pp}})}{dxdQ^2}\right) {\cal A}(E_\ell,\theta_\ell,E_h)  \, . 
\ee
Given that the DIS cross-section depends linearly on the PDFs, Eq.~(\ref{eq:event_yields_calculation_nuPDF}) has the same formal structure of a hadron-hadron collision, where the partonic cross-section is convoluted with two PDFs.
Note only that, as opposed to quark and gluon PDFs, the neutrino PDF defined in this way is scale-independent.\footnote{The scale independence of the neutrino PDFs defined here would be broken were one to account for electroweak corrections~\cite{Bauer:2017isx} which are neglected in this work.} 

\paragraph{Implementation.}
The {\sc\small POWHEG-BOX-RES} neutrino DIS code of~\cite{FerrarioRavasio:2024kem} allows to switch between neutrinos of fixed energy and neutrinos with a broadband energy distribution.
We exploit the latter possibility to interface the neutrino PDFs constructed in Eq.~(\ref{eq:neutrino_pdf_definition}) in {\sc\small POWHEG} by means of the corresponding {\sc\small LHAPDF} grid file.
Varying experiment and (anti)neutrino flavour is possible by selecting the associated grid name and PDG ID number, respectively within the {\sc\small POWHEG} run card.
Specifically, the user needs to set \texttt{fixed\_lepton\_beam} to zero and \texttt{nupdf} to the {\sc\small LHAPDF} ID of the neutrino PDF. 
The energy of the (anti)neutrino beam should be set to the centre-of-mass energy of the pp collision, $\sqrt{s_{\rm pp}}$, in a consistent manner with the definition of Eq.~(\ref{eq:xnu}).

The neutrino fluxes for the LHC far-forward neutrino experiments considered in this work are taken from the calculations of~\cite{Kling:2021gos}, except for the contribution from $D$-meson decays which instead is taken from the recent analysis of~\cite{Buonocore:2023kna}, based on {\sc\small POWHEG} NLO simulations with the NNPDF3.1 BFKL-resummed variant of~\cite{NNPDF:2017mvq, Ball:2017otu} as input PDF.
Neutrino fluxes are normalised according to Eq.~(\ref{eq:neutrino_pdf_definition}) and stored in interpolation grids complying with the {\sc\small LHAPDF} interface format, after applying a smoothing function to facilitate the correct numerical behaviour of the {\sc\small LHAPDF} interpolator. 
These grids have support for $0\le x_{\nu}\le 1$ and the meta-data is set up in a way such that the output is independent on the choice of $Q^2$.

The resulting {\sc\small LHAPDF} grid file provides the values of the neutrino PDFs,
\be
\label{eq:nuPDFs_lhapdf_grid}
f^{(k)}_{\nu_e}(x_\nu),~f^{(k)}_{\nu_\mu}(x_\nu),~f^{(k)}_{\nu_\tau}(x_\nu),~
f^{(k)}_{\bar{\nu}_e}(x_\nu),~f^{(k)}_{\bar{\nu}_\mu}(x_\nu),~f^{(k)}_{\bar{\nu}_\tau}(x_\nu) \, ,\quad k=0,\ldots,{N_{\rm pdf}-1} \, ,
\ee
for any value of the neutrino momentum
fraction $x_\nu$ defined in Eq.~\eqref{eq:xnu}.
Each (anti-)neutrino flavour is accessed via its corresponding PDG number. 
Here $N_{\rm pdf}$ stands for the number of members of this neutrino PDF set.
These members may correspond to different experiments (for which one has a different flux) but also to alternative calculations of a given flux for the same experiment, for example varying the theory settings (e.g. the scales $\mu_F$ and $\mu_R$ for charm production) or using alternative MC generators to compute the LHC forward hadron fluxes.
The chosen format is agnostic with respect to these choices, which should be spelled out in the meta-data of the {\sc\small LHAPDF} grid file.
One may also choose to generate different grid files, one for each experiment: ultimately, the best choice depends on the specific applications being sought.\footnote{The neutrino PDF grids released with this paper can be accessed using the following {\sc\small LHAPDF} IDs: 40011300 (FASER$\nu$), 40011500 (SND@LHC), 
40011700 (FASER$\nu$2), and 40011900 (FLArE).}

Table~\ref{tab:FPF_experiments} summarises the main settings of the LHC far-forward neutrino experiments that will be considered in this work and for which neutrino PDFs are constructed. 
Since modelling tau-neutrino reconstruction requires detector simulation, here for simplicity we apply the same acceptance cuts as for muon neutrinos.
For FLArE, we assume that muons would be measured in the FASER2 spectrometer situated downstream in the FPF cavern.
The neutrino fluxes used assume that FASER$\nu$ and SND@LHC acquire data for the LHC Run-3 period ($\mathcal{L}=150$ fb$^{-1}$), while FASER$\nu$2 and FLArE take data for the complete HL-LHC period ($\mathcal{L}=3$ ab$^{-1}$).

\begin{table*}[t]
  \centering
  \small
  \renewcommand{\arraystretch}{1.40}
\begin{tabularx}{\textwidth}{Xccccc}
\toprule
Detector &  Rapidity range &  Target material & Kinematic acceptance  \\
\midrule
\midrule
\multirow{2}{*}{FASER$\nu$}  &  \multirow{2}{*}{ $\eta_\nu \ge 8.5$}  &   \multirow{1}{*}{Tungsten}       &   $E_\ell,E_h \gsim 100$ GeV     \\
&   &   \multirow{1}{*}{(1.1 tonnes)}  &        $\tan \theta_\ell \lsim 0.025 $          \\
\midrule
\multirow{2}{*}{SND@LHC}  & \multirow{2}{*}{ $7.2 \le \eta_\nu \le 8.4$}   &  Tungsten   &     $E_\ell,E_h \gsim 20 $ GeV           \\
  &    &  (0.83 tonnes)   &    $\theta_\mu \lsim 0.15, \theta_e \lsim 0.5$               \\
\midrule
\midrule
\multirow{2}{*}{FASER$\nu$2}  & \multirow{2}{*}{ $\eta_\nu \ge 8.5$}  & \multirow{1}{*}{Tungsten}    &    $E_\ell,E_h \gsim 100$ GeV      \\
  &   &  \multirow{1}{*}{(20 tonnes)}         &  $\tan \theta_\ell \lsim 0.05$   \\
\midrule
\multirow{2}{*}{FLArE}  & \multirow{2}{*}{$\eta_\nu \ge 7.5$} & LAr  & $E_\ell,E_h \gsim 2$ GeV, $E_e \lsim 2$ TeV     \\
&   &  (10~tonnes)   &   $\theta_\mu \lsim 0.025$, $\theta_e \lsim 0.5$  \\
  \bottomrule
\end{tabularx}
\vspace{0.2cm}
\caption{\small Overview of the neutrino pseudo-rapidity coverage, target material,
and final-state kinematic acceptance of the four far-forward LHC neutrino experiments considered in this work.
Whenever required, we indicate separately the acceptance for electrons (positrons) and (anti-)muons. 
  The neutrino fluxes used assume that FASER$\nu$ and SND@LHC acquire data
  for the LHC Run-3 period ($\mathcal{L}=150$ fb$^{-1}$), while FASER$\nu$2 and and FLArE take data
  for the complete HL-LHC run ($\mathcal{L}=3$ ab$^{-1}$).
  See~\cite{Cruz-Martinez:2023sdv} for more details. 
  \label{tab:FPF_experiments}
}
\end{table*}

\begin{figure}[t]
\centering
\includegraphics[width=0.99\linewidth]{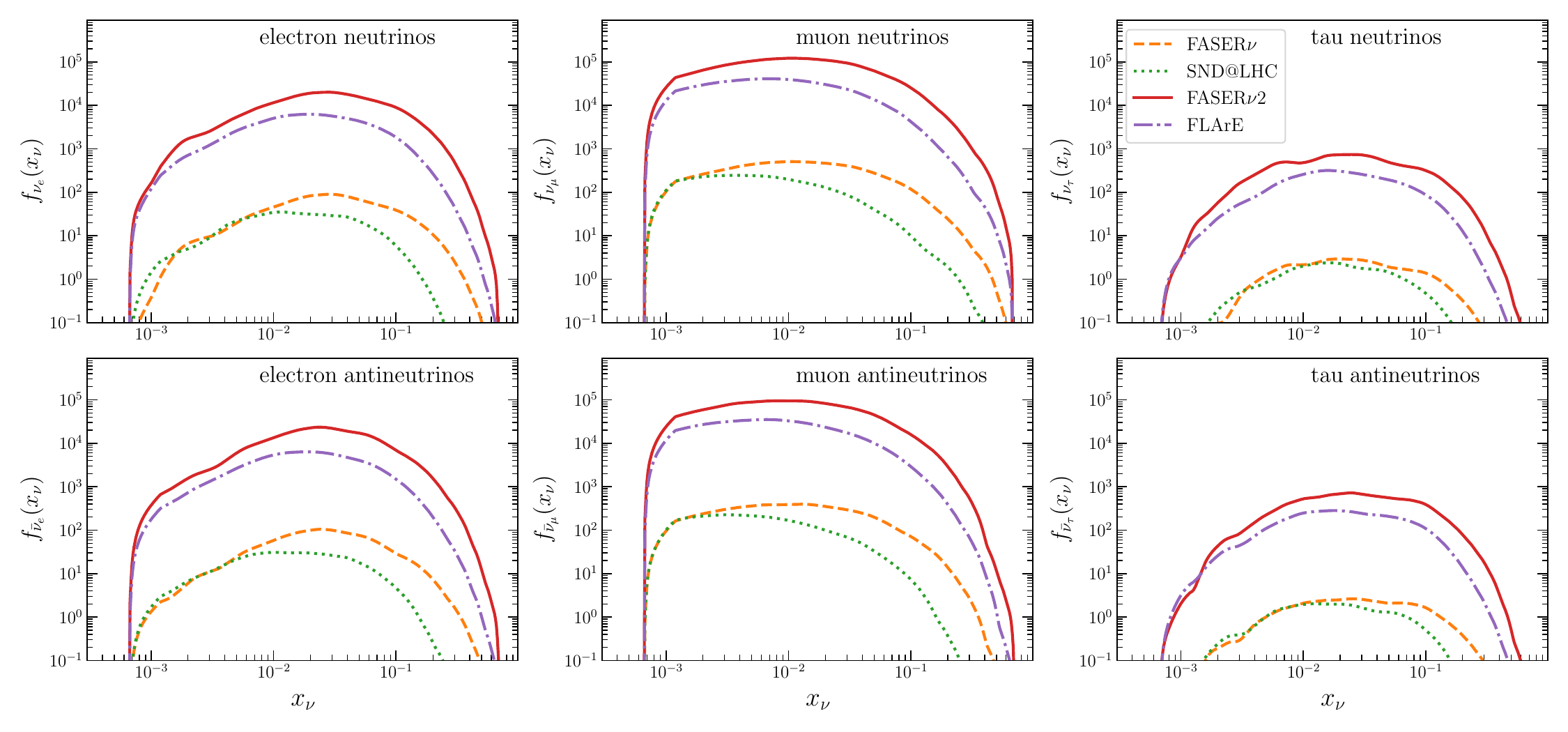}
\caption{The neutrino PDFs $f_{\nu_i}(x_\nu)$ for the four LHC far-forward experiments considered in this work. 
From left to right, we display in the top (bottom) panels the (anti-)neutrino PDFs for electron, muon, and tau neutrinos. 
}
\label{fig:neutrino-pdfs-exps}
\end{figure}

\begin{figure}[t]
\centering
\includegraphics[width=0.89\linewidth]{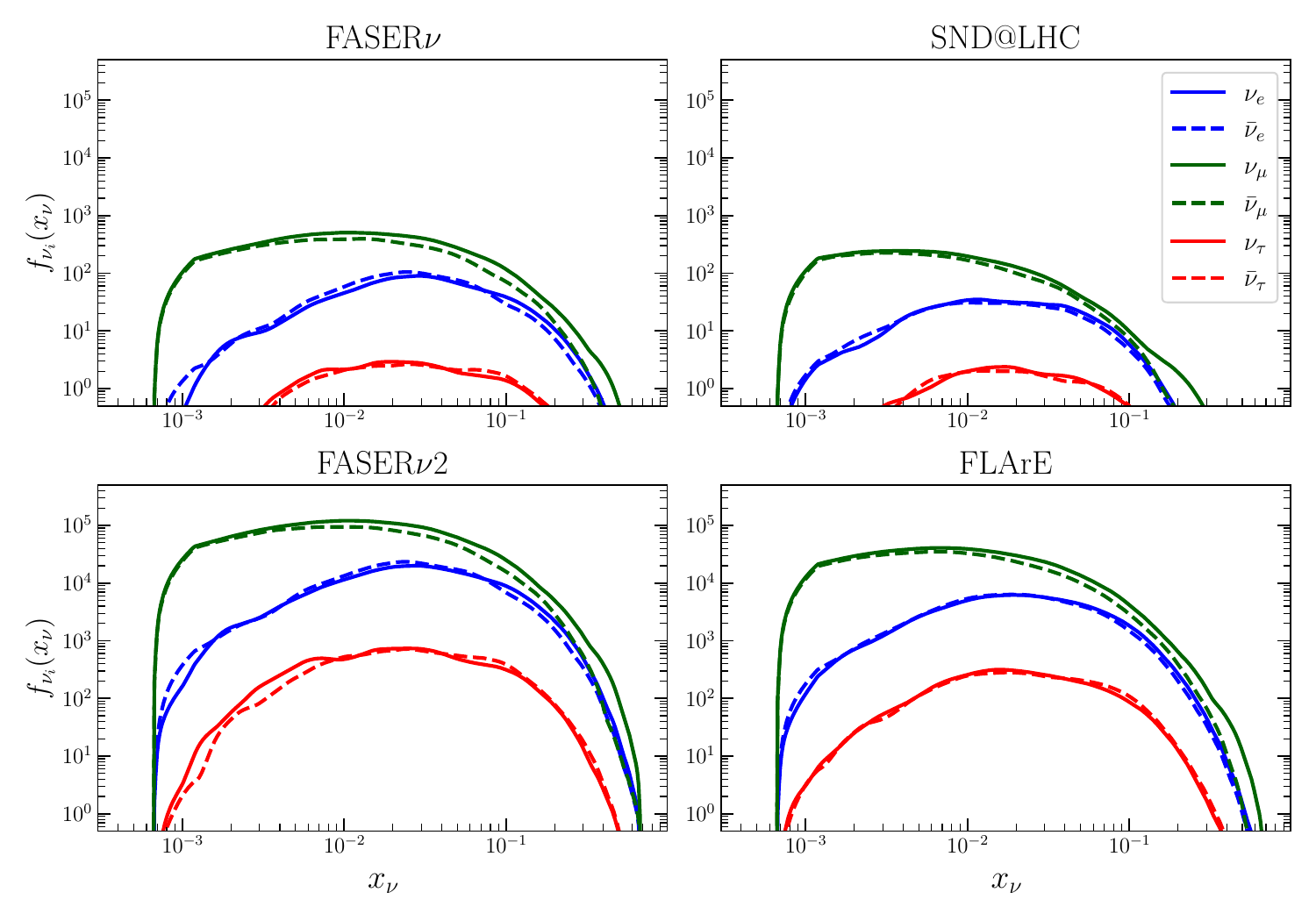}
\caption{Same as Fig.~\ref{fig:neutrino-pdfs-exps} comparing the neutrino and antineutrino PDFs of different flavours for the FASER$\nu$, SND@LHC, FASER$\nu$2, and FLArE experiments.
}
\label{fig:neutrino-pdfs-flavour}
\end{figure}

In Figs.~\ref{fig:neutrino-pdfs-exps} and~\ref{fig:neutrino-pdfs-flavour} we display the neutrino PDFs for the four LHC far-forward experiments considered in this work: FASER$\nu$ and SND@LHC (Run-3) and FASER$\nu$2 and FLArE (HL-LHC).
One observes how these neutrino PDFs display a relatively broad distribution in $x_\nu$.
The electron neutrino fluxes
peak around $x_\nu\simeq 0.05$, corresponding to neutrinos with energy around $E_\nu\sim 700$ GeV, while the muon neutrino fluxes exhibit a plateau for $x_\nu\in \lc 10^{-3},0.1\rc$. 
All fluxes show a steep fall-off in the small- and large-$x_\nu$ regions.
While muon neutrinos are associated with the largest overall fluxes, at large $x_\nu$, the size of the electron-neutrino PDFs becomes comparable. 
Small asymmetries between neutrino and antineutrino fluxes arise from differences in the underlying production mechanisms~\cite{Kling:2021gos,Buonocore:2023kna}.
Figs.~\ref{fig:neutrino-pdfs-exps} and~\ref{fig:neutrino-pdfs-flavour}  highlight the much higher rates that are expected at the HL-LHC far-forward experiments as compared to the Run-3 ones.

\begin{table}[t]
    \centering
    \small
 \renewcommand{\arraystretch}{1.90}
 \begin{tabularx}{\textwidth}{|X|cc|cc|}
    \Xhline{3\arrayrulewidth} 
         \multirow{2}{*}{Experiment } & \multicolumn{2}{c|}{before cuts} & \multicolumn{2}{c|}{after cuts} \\
          & \multicolumn{1}{l}{$N^{(\nu_e)}_{\rm int} +N^{(\bar{\nu}_e)}_{\rm int}$} & \multicolumn{1}{l|}{$N^{(\nu_\mu)}_{\rm int} +N^{(\bar{\nu}_\mu)}_{\rm int}$} &  \multicolumn{1}{l}{$N^{(\nu _e)}_{\rm int} +N^{(\bar{\nu}_e)}_{\rm int}$} & \multicolumn{1}{l|}{$N^{(\nu _\mu)}_{\rm int} +N^{(\bar{\nu}_\mu)}_{\rm int}$} \\
          \cline{1-5}
         FASER$\nu$ & $1109_{-1.1\%}^{+0.8\%}$ & $3783_{-1.2\%}^{+1.0\%}$ & $554_{-1.3\%}^{+1.2\%}$ & $1613_{-1.5\%}^{+1.3\%}$  \\
         SND@LHC  & $263_{-1.3\%}^{+1.1\%}$ & $836_{-1.4\%}^{+1.4\%}$ & $240_{-1.3\%}^{+1.1\%}$ & $679_{-1.5\%}^{+1.2\%}$ \\
    \cline{1-5}

          FASER$\nu$2 & $(250\times10^3)_{-1.2\%}^{+0.8\%}$ & $(930\times10^3)_{-1.2\%}^{+1.0\%}$ & $(172\times10^3)_{-1.0\%}^{+0.7\%}$ & $(526\times10^3)_{-1.0\%}^{+0.7\%}$\\

         FLArE & $(62\times10^3)_{-1.2\%}^{+1.1\%}$ & $(219\times10^3)_{-1.3\%}^{+1.2\%}$ & $(60\times10^3)_{-1.2\%}^{+1.1\%}$ & $(110\times10^3)_{-1.8\%}^{+1.0\%}$ \\
    \Xhline{3\arrayrulewidth}
    \end{tabularx}
    \vspace{0.3cm}
    \caption{The number of interacting neutrino events evaluated with \POWHEG~NLO and neutrino PDFs using  Eq.~(\ref{eq:event_yields_calculation_nuPDF}), for the four LHC far-forward experiments considered in this work. 
      We indicate the total number of electron and muon neutrino scattering events expected both before and after imposing the acceptance cuts of Table~\ref{tab:FPF_experiments}.
      Theoretical uncertainties from scale variations are evaluated using the 7-point prescription. 
      The settings adopted to obtain these \POWHEG~event yields are described in Sect.~\ref{sec:results}. 
    }
    \label{tab:events_after_cuts}
\end{table}

To illustrate the impact of acceptance selection cuts on the total event yields, Table~\ref{tab:events_after_cuts} indicates the number of interacting neutrino events evaluated at NLO with {\sc\small POWHEG} and neutrino PDFs using  Eq.~(\ref{eq:event_yields_calculation_nuPDF}) for the four experiments considered, adopting the settings to be described in Sect.~\ref{sec:results}. 
We indicate the total number of electron and muon neutrino scattering events expected both before and after imposing the acceptance cuts from in Table~\ref{tab:FPF_experiments}.
Consistently with~\cite{Cruz-Martinez:2023sdv}, for the FASER$\nu$ and FASER$\nu$2 experiments the acceptance cuts reduce the yields by about 40\%, while for  SND@LHC and FLArE acceptance cuts do not markedly modify the predicted yields. 

To validate the implementation of the neutrino PDF formalism in {\sc\small POWHEG}, we have computed the NLO inclusive event yields Eq.~(\ref{eq:event_yields_calculation}) for
the four experiments considered, and compared the results with those in Table~2.2 of~\cite{Cruz-Martinez:2023sdv}.
The latter was based on evaluating Eq.~(\ref{eq:event_yields_calculation})  using NLO neutrino structure functions computed with {\sc\small YADISM} with PDF4LHC21 NNLO~\cite{PDF4LHCWorkingGroup:2022cjn} as input PDF.
Both calculations are restricted to the DIS  kinematic region defined by $Q^2 \ge 2$ GeV$^2$ and $W \ge 2$ GeV.
The {\sc\small POWHEG}-based predictions for the inclusive event yields are found to be
qualitatively consistent with the  independent calculation of \cite{Cruz-Martinez:2023sdv} based on {\sc\small YADISM}, further validating the implementation of the neutrino PDF formalism in {\sc\small POWHEG}. \footnote{ To be specific, the ``before cuts'' event yields in Table~\ref{tab:events_after_cuts} are the same quantity, with different input PDFs and calculational settings, than the ``before cuts'' yields in Table~2.2 of~\cite{Cruz-Martinez:2023sdv}, while the ``after cuts'' yields here can be compared to the ``after DIS and acceptance cuts'' yields in Table~2.2 of~\cite{Cruz-Martinez:2023sdv}. }

\section{The fixed neutrino energy case}
\label{sec:results_validation}

We here present results for {\sc\small POWHEG} NLO simulations of neutrino-proton scattering for fixed neutrino energy $E_\nu$.
We perform a tuned comparison of  {\sc\small POWHEG} LO calculations with  stand-alone {\sc\small Pythia8} simulations for differential distributions.
For the same observables, we evaluate the ratio of the {\sc\small POWHEG} NLO predictions to the LO ones, quantifying the impact of QCD corrections.
We also validate {\sc\small POWHEG}  NLO calculations for double-differential DIS cross-sections with those provided by the analytic {\sc\small YADISM} framework based on the structure function formalism.
We assess the interplay between NLO QCD corrections and final-state acceptance selections.
We also compare the {\sc\small POWHEG} NLO predictions for final-state distributions with their counterparts evaluated with the {\sc\small GENIE} neutrino event generator.

\begin{figure}[t]
    \begin{minipage}[t]{0.49\linewidth}
        \includegraphics[width=\linewidth]{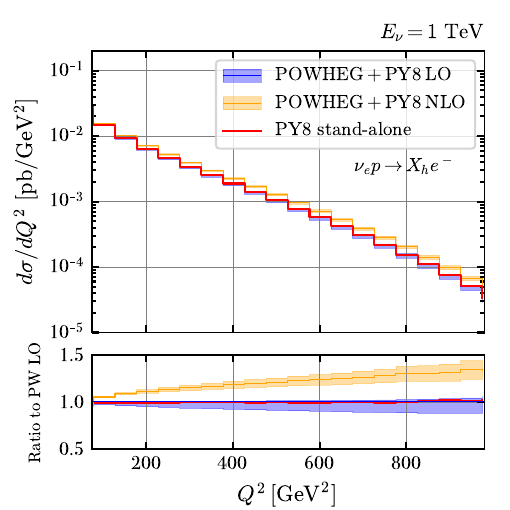}
    \end{minipage}
    \begin{minipage}[t]{0.49\linewidth}
        \includegraphics[width=\linewidth]{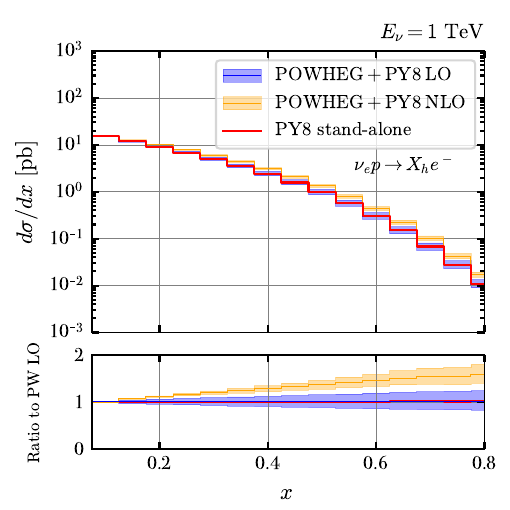}
    \end{minipage}
    \caption{The single-differential distributions in $Q^2$ (left) and Bjorken-$x$ (right) in charged-current electron neutrino-proton scattering with $E_\nu=1$ TeV. 
    We compare the stand-alone {\sc\small Pythia8} calculation with its counterparts based on {\sc\small POWHEG+Pythia8} both at LO and at NLO accuracy, in both cases with error bands accounting for the MHOUs.
    The only kinematic cuts applied are 
    $Q^2 \ge 10$ GeV$^2$ and
$W \ge 2$ GeV.
    The bottom panels display the ratio with respect to the central value of the {\sc\small POWHEG} (PW) LO calculation.
    }    
    \label{fig:initial_variables}
\end{figure}

\begin{figure}[t]
    \begin{minipage}[t]{0.33\linewidth}
        \includegraphics[width=\linewidth]{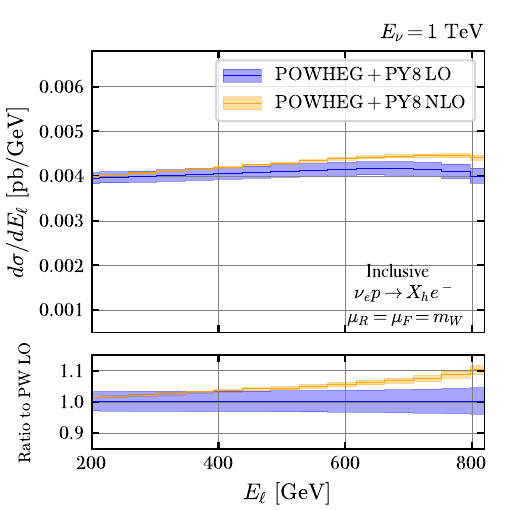} 
    \end{minipage}
    \begin{minipage}[t]{0.33\linewidth}
        \includegraphics[width=\linewidth]{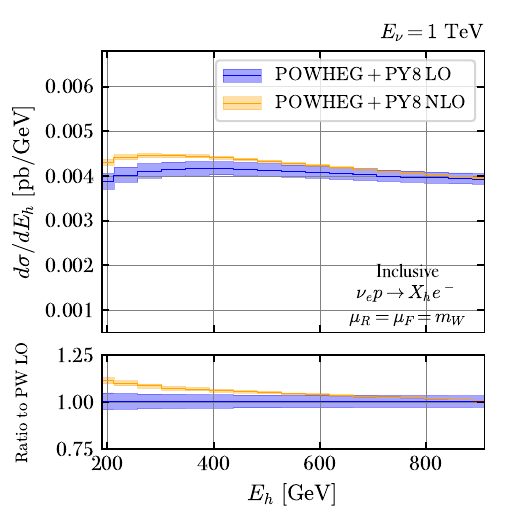} 
    \end{minipage}
    \begin{minipage}[t]{0.33\linewidth}
        \includegraphics[width=\linewidth]{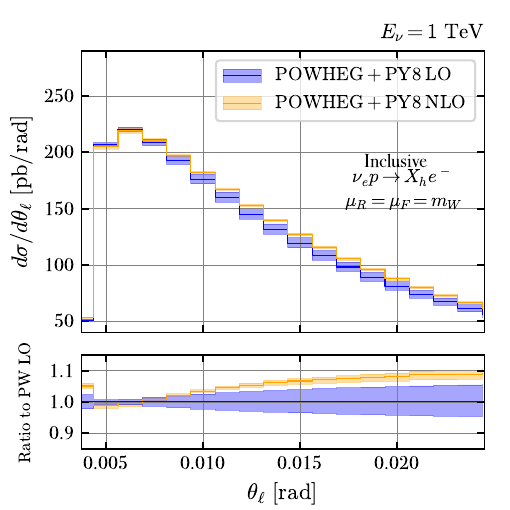}
    \end{minipage}
    \vspace{2em} 
    \begin{minipage}[t]{0.33\linewidth}
        \centering
        \includegraphics[trim={0cm, 0cm, 0cm, 0.725cm}, clip, width=\linewidth]{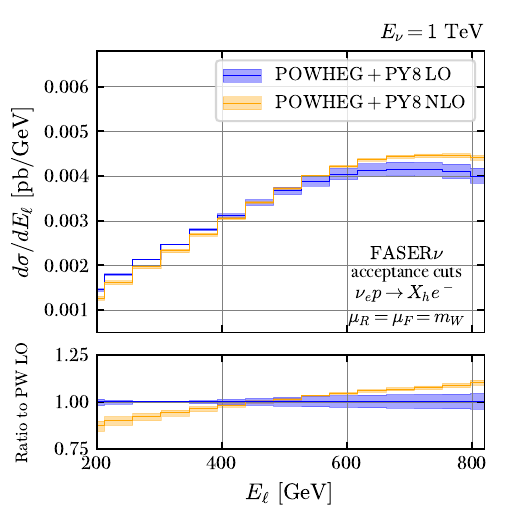}
    \end{minipage}
    \begin{minipage}[t]{0.33\linewidth}
        \centering
        \includegraphics[trim={0cm, 0cm, 0cm, 0.725cm}, clip, width=\linewidth]{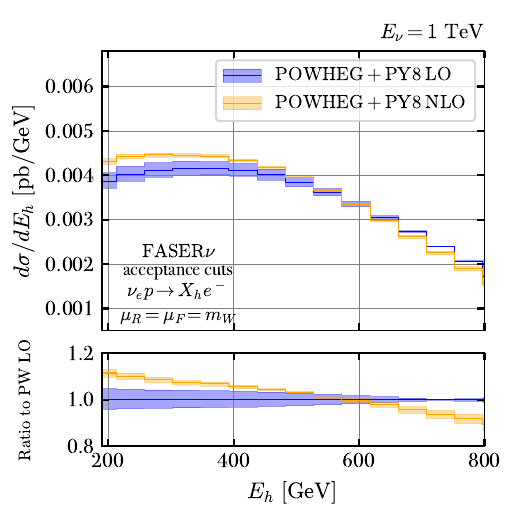} 
    \end{minipage}
    \begin{minipage}[t]{0.33\linewidth}
        \centering
        \includegraphics[trim={0cm, 0cm, 0cm, 0.725cm}, clip, width=\linewidth]{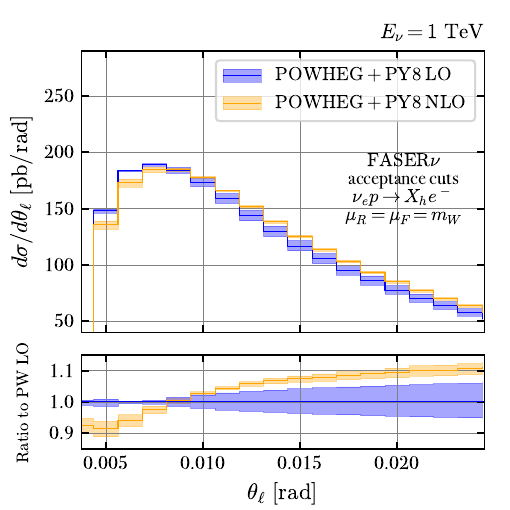}
    \end{minipage}
    \vspace{-1cm}
    \caption{Same as Fig.~\ref{fig:initial_variables}
    for the single-differential distributions in the lepton energy $E_\ell$, the 
    hadronic final-state energy $E_h$, and the lepton scattering angle $\theta_\ell$. 
    We present results both inclusive in the final-state kinematics (top) and for the case of the FASER$\nu$ acceptance cuts listed in Table~\ref{tab:FPF_experiments} (bottom).
    Scales are set to $\mu_F=\mu_R=m_W$.
   The {\sc\small POWHEG+Pythia8} LO predictions coincide with the {\sc\small Pythia8} stand-alone ones (not shown here).
    }
    \label{fig:final_variables}
\end{figure}

\paragraph{Comparison with {\sc\small Pythia8} stand-alone and impact of NLO corrections.}
The input settings such as PDFs, CKM matrix elements, electroweak couplings, and vector boson masses are taken to be:
\begin{align}
   & G_F=1.16637\times 10^{-5}~{\rm GeV^{-2}} \, , \quad
    m_W=80.385~{\rm GeV} \,, \quad
    m_Z=91.1876~{\rm GeV} \, ,\nonumber \\
  &  |V_{ud}| = 0.97383 \,, \quad \label{eq:input_parameters}
    |V_{us}| = 0.2272 \, ,\quad 
    |V_{cd}| = 0.2271  \, , \\ \nonumber    & |V_{cs}| = 0.975\,, \quad
     \, 
     |V_{cb}| = 42.21 \times 10^{-3}\,, \quad
 |V_{ub}| = 3.96 \times 10^{-3}\,, \\ \nonumber
&|V_{td}| = 8.14 \times 10^{-3} \,, \quad
|V_{ts}| = 41.61 \times 10^{-3}\,, \quad
|V_{tb}| = 0.9991 \,.
\end{align}
At leading order in the electroweak coupling, the Fermi constant $G_F$ is related
to the electroweak coupling constant
$g_{\rm EW}$ and to the QED coupling
$\alpha_{\rm QED}$ by 
\be
G_F = \frac{\sqrt{2}}{8}
\frac{g_{\rm EW}^2}{m_W^2}= \frac{\pi \alpha_{\rm QED}}{\sqrt{2}m_W^2(1-m_W^2/m_Z^2)} \, .
\ee
We do not consider higher-order QED or electroweak corrections, and hence the electroweak couplings are scale independent. 
The input PDF is the central replica of the NNPDF4.0 NNLO~set\cite{NNPDF:2021njg,NNPDF:2021uiq}.
The renormalisation and factorisation scales are set dynamically to be $\mu_F=\mu_R=Q$, except for Figs.~\ref{fig:initial_variables} and~\ref{fig:final_variables} for which we adopt a fixed scale $\mu_F=\mu_R=m_W$.
The value of the QCD coupling constant and its running are taken from NNPDF4.0 and given by $\alpha_s(m_Z)=0.1180$, consistent with the latest PDG average~\cite{ParticleDataGroup:2022pth}.
The Monash 2013 tune of {\sc\small Pythia8}~\cite{Skands:2014pea} is used, and in Sect.~\ref{sec:results} we compare it with the forward tune of~\cite{Fieg:2023kld}. 

As discussed in~\cite{FerrarioRavasio:2024kem}, the {\sc\small POWHEG-BOX-RES} implementation of DIS processes assumes massless projectiles.
To reproduce the fixed-target (proton at rest) kinematics, it suffices to run the code with the proton energy set to $E_p =m_p/2$ and the neutrino energy to $\widetilde{E}_\nu = E_\nu + m_p/2$.
This way, the collision centre of mass energy, given by $\sqrt{s}=2\sqrt{E_\nu E_p}$, corresponds to the same quantity evaluated in the target rest frame, $\sqrt{s}=\sqrt{m_p\lp 2E_\nu + m_p\rp}$, up to  $m_p/E_\nu$ corrections which are small for the LHC neutrino kinematics. 
Selecting this fixed-target option is possible through the {\sc\small POWHEG} run card.

Fig.~\ref{fig:initial_variables} displays the single-differential cross-sections in $Q^2$ and Bjorken-$x$ in charged-current neutrino-proton scattering for $E_\nu=1$ TeV. 
We compare the stand-alone {\sc\small Pythia8} calculations with those obtained from {\sc\small POWHEG+Pythia8}.
The {\sc\small POWHEG} predictions include an estimate of the missing higher order uncertainties (MHOUs) evaluated by means of the 7-point prescription for scale variations.
The only kinematic cuts applied are 
$Q^2 \ge 10$ GeV$^2$ and $W \ge 2$ GeV, while no restrictions on the final-state acceptance are applied.
The bottom panels of Fig.~\ref{fig:initial_variables} display the ratio to the central value of the {\sc\small POWHEG} LO calculation.
We verify that the stand-alone {\sc\small Pythia8} predictions reproduce those from {\sc\small POWHEG} LO.
Integrating the {\sc\small POWHEG} distributions over either $x$ or $Q^2$, we recover the  total neutrino inclusive cross-sections reported in~\cite{Bertone:2018dse} and exhibiting an overall negative NLO $K$-factor of a few percent. 
While the $K$-factors shown in Fig.~\ref{fig:initial_variables} are always positive, the inclusive cross-section is dominated by the small-$Q^2$ (with $Q^2>10$ GeV$^2$) and $x$ region, not shown here.
In that region, the $K$-factor is negative and  hence also that of the total cross-section. 

From Fig.~\ref{fig:initial_variables} one observes that for neutrinos with energy of $E_\nu=1$ TeV, NLO QCD corrections to the single-differential distribution in $x$ ($Q^2$) increase with the partonic momentum fraction (momentum transfer), reaching +60\% at $x=0.8$ (+40\% at $Q^2=10^3$ GeV$^2$).
Qualitatively similar results are found for the antineutrino distributions.
Hence, while NLO QCD corrections to CC neutrino DIS in the TeV region are moderate (few-percent level) for total inclusive cross-sections, in general they become more significant at the level of differential distributions 

\paragraph{Impact of acceptance cuts.}
Fig.~\ref{fig:final_variables} displays {\sc\small POWHEG+Pythia8} LO and NLO predictions for the single-differential distributions in the final-state variables $E_\ell$ (charged-lepton energy), $E_h$ (energy of hadronic final state), and $\theta_\ell$ (charged-lepton scattering angle). 
We provide results both without any acceptance cuts and for the case in which the FASER$\nu$ acceptance cuts listed in Table~\ref{tab:FPF_experiments} are applied. 
The stand-alone {\sc\small Pythia8} calculation, not shown here, reproduces the {\sc\small POWHEG} LO results in all cases.

From the bottom panels of Fig.~\ref{fig:final_variables} one observes that, in the case of final-state observables within the FASER$\nu$ acceptance,   NLO QCD corrections
depend on the kinematics.
The QCD $K$-factor ranges between $-15\%$ and $+8\%$ for the $E_\ell$ distribution and between $-10\%$ and $+10\%$ for the $E_h$ and $\theta_\ell$ distributions. 
The MHOUs associated to the NLO predictions are typically a few percent, with mild dependence on the kinematics.
Imposing acceptance cuts modifies the shapes of the three distributions, specially for $E_\ell$ and $E_h$, and as a consequence the pattern of NLO QCD corrections is rather different at the inclusive and fiducial levels. 
For instance, while for the $E_\ell$ 
and $E_h$ distributions the NLO corrections are positive in the considered phase space for the inclusive selection, they can become negative once the FASER$\nu$ acceptance cuts are applied.
These results emphasise the relevance of NLO event generators for the modelling of differential distributions in the presence of realistic acceptance cuts.

\begin{figure}[t]
    \centering
\includegraphics[width=0.49\textwidth]{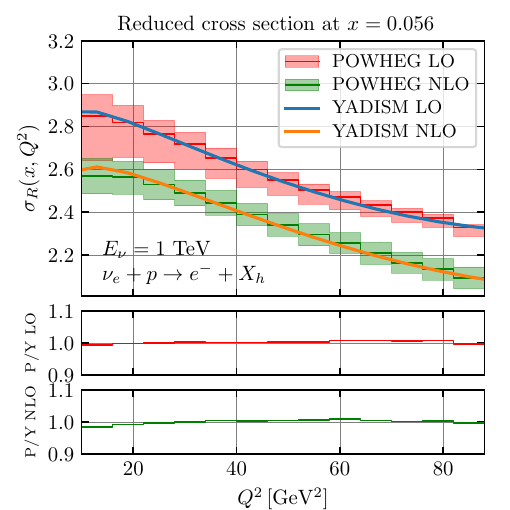}
\includegraphics[width=0.49\textwidth]{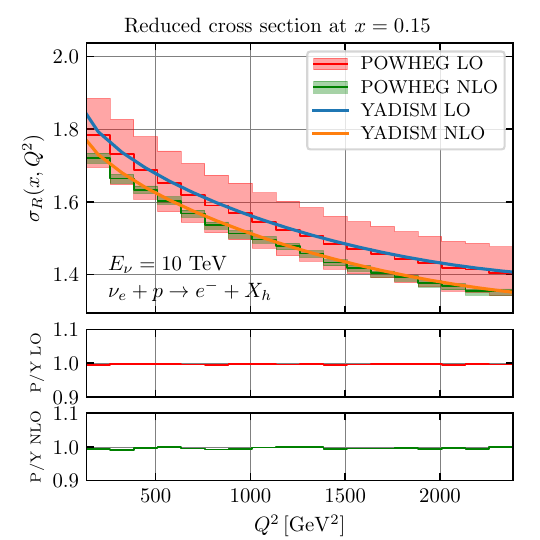}
    \caption{The double-differential DIS reduced cross-section, 
    $\sigma^{\nu p}_{R}(x,Q^2,E_\nu)$, evaluated for  
    neutrino-proton scattering at $x = 0.056$ and $E_\nu=1~{\rm TeV}$ (left) and  $x = 0.15$ and $E_\nu=10~{\rm TeV}$ (right panel) and  as a function of $Q^2$. 
    The {\sc\small POWHEG} cross section is binned in $x$ and $Q^2$ with a bin width of $\Delta x = 0.02$.
    We compare the {\sc\small POWHEG} LO and NLO calculations (without matching to {\sc\small Pythia8}) with their
   {\sc\small YADISM} counterparts.
    The bottom panels display the ratios between the {\sc\small POWHEG}
    and {\sc\small YADISM} results for the two perturbative orders.
    }
    \label{fig:sigmaR_CC_x056}
\end{figure}

\paragraph{Comparison with {\sc\small YADISM}.}
The {\sc\small POWHEG} implementation of neutrino DIS reproduces the {\sc\small YADISM}~\cite{Candido:2024rkr} predictions for double-differential inclusive cross-sections based on the structure function formalism.
Fig.~\ref{fig:sigmaR_CC_x056} displays the DIS reduced cross-section $\sigma^{\nu p}_{R}(x,Q^2,E_\nu)$, Eq.~(\ref{eq:sigmaR_CC}), evaluated for neutrino-proton scattering at  $x = 0.056$ and $E_\nu=1~{\rm TeV}$ and at $x = 0.15$ and $E_\nu=10~{\rm TeV}$ (left and right panels, respectively) as a function of the momentum transfer $Q^2$.
The {\sc\small POWHEG} cross section is binned in $x$ and $Q^2$ with a bin width of $\Delta x = 0.02$, and MHOUs are only displayed for the {\sc\small POWHEG} predictions. 
Here the {\sc\small POWHEG} calculation is not matched to {\sc\small Pythia8}, given that no acceptance cuts to the final-state are applied to enable the consistent comparison with the structure function calculation of {\sc\small YADISM}.
Good agreement is found between the {\sc\small POWHEG} and {\sc\small YADISM} predictions, both at LO and at NLO.\footnote{In the large-$x$ and small-$Q^2$ regions, differences between the two codes may arise due to the treatment of $m_p\ne 0$ effects.}

\begin{figure}[t]
\includegraphics[width=0.33\linewidth]{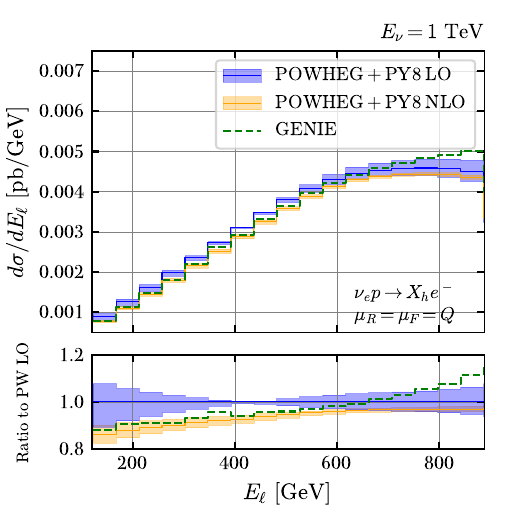} 
\includegraphics[width=0.33\linewidth]{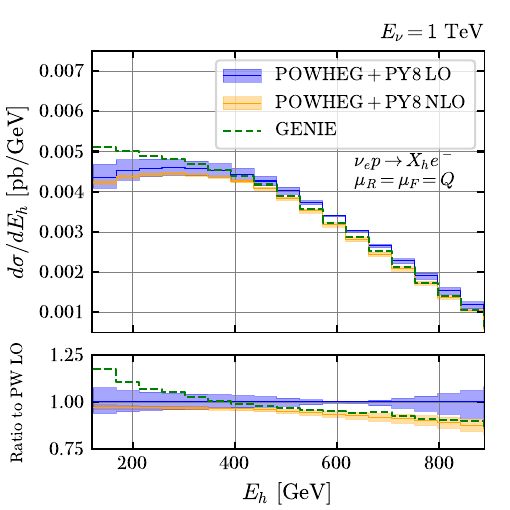} 
\includegraphics[width=0.33\linewidth]{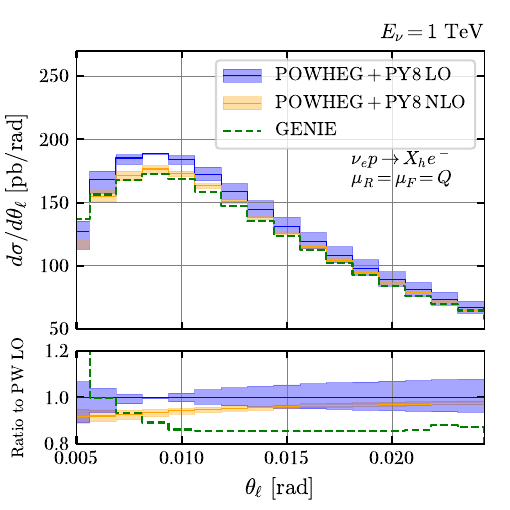} 
\caption{Same as     Fig.~\ref{fig:final_variables} for the $E_\ell$, $E_h$, and $\theta_\ell$ distributions within the FASER$\nu$ acceptance for $E_\nu = 1~\mathrm{TeV}$, comparing the outcome of {\sc\small POWHEG+Pythia8} LO and NLO simulations with those from  {\sc\small GENIE}.
The latter are based the HEDIS tune with structure functions evaluated with HERAPDF1.5~\cite{Cooper-Sarkar:2010yul}.
Note that, in contrast to Fig.~\ref{fig:final_variables}, here we use a dynamical scale choice of $\mu_F=\mu_R=Q$ in both calculations.
}
\label{fig:final_variables_genie}
\end{figure}

\paragraph{Comparison with {\sc\small GENIE}.}
We compare in Fig.~\ref{fig:final_variables_genie} the output of {\sc\small POWHEG+Pythia8} LO and NLO simulations for the $E_\ell$, $E_h$, and $\theta_\ell$ final-state distributions in $\nu_e+p$ scattering within the FASER$\nu$ acceptance at $E_\nu=1$ TeV with those from {\sc\small GENIE}.  
The latter are based on one of the HEDIS tunes~\cite{Garcia:2020jwr}, with structure functions evaluated with HERAPDF1.5 NLO~\cite{Cooper-Sarkar:2010yul} as input PDF set.
As opposed to the previous (tuned) comparisons, in Fig.~\ref{fig:final_variables_genie} one expects differences between the {\sc\small GENIE} and {\sc\small POWHEG+Pythia8} results, given that the  hard-scattering matrix elements, PDFs, parton showers, and other input settings are disparate in the two calculations.

From Fig.~\ref{fig:final_variables_genie} 
we observe differences up to 20\%, depending on the specific distribution and the kinematic region, with either a positive or a negative shift. 
For instance, for the $E_h$ distribution, {\sc\small GENIE} predicts a higher cross-section than {\sc\small POWHEG+Pythia8} NLO by an amount of around 20\% in the low $E_h$ region.
Differences of this magnitude are qualitatively consistent with those found in previous studies~\cite{Candido:2023utz,FASER:2024ykc}.
For the low-$E_\ell$ and high-$E_h$ distributions, good agreement is found between  {\sc\small GENIE} and the {\sc\small POWHEG+Pythia8} NLO calculations. 
The worse disagreement is observed for the $\theta_\ell$ distribution.

The comparisons in Fig.~\ref{fig:final_variables_genie} illustrate the main advantages of using {\sc\small POWHEG} instead of {\sc\small GENIE} for simulations at the LHC far-forward neutrino experiments: higher perturbative accuracy, robust estimate of theoretical uncertainties, state-of-the-art PDFs, and availability of modern parton shower and hadronisation models.
These considerations also imply that taking the differences between {\sc\small GENIE} and {\sc\small POWHEG} to estimate the underlying theory error in the calculation would be overly conservative, due to the mismatch in their perturbative accuracy.

\section{NLO+PS predictions for LHC far-forward neutrino experiments}
\label{sec:results}

We now present {\sc\small POWHEG} predictions accurate at NLO+PS for differential distributions within  acceptance for ongoing (FASER$\nu$ and SND@LHC) and future (FASER$\nu$2 and FLArE) far-forward LHC neutrino experiments.
We quantify the impact of NLO QCD corrections and study the stability of the results with respect to variations of the parton shower and of the choice of the {\sc\small Pythia8} tune.
We also study the dependence of our results with respect to the incoming neutrino flavour, and evaluate cross-section ratios between neutrino flavours which are relevant for phenomenological applications.
First we focus on predictions for the FASER$\nu$ detector, and then we assess how results vary for other experiments.

In the following we adopt the following numerical values of the input parameters:
\bea
   && G_F=1.1663788\times 10^{-5}~{\rm GeV^{-2}} \, , 
    \alpha_{\rm em}=7.297\times 10^{-3} \, , \quad 
    \sin^2 \theta_W=0.23121 \, , \quad\nonumber\\[0.1cm]
    &&   m_W=80.377~{\rm GeV} \,, \quad
 \Gamma_W=2.085~{\rm GeV}\,, \quad   \nonumber \\
   &&  m_Z=91.1876~{\rm GeV} \, ,\quad
   \Gamma_Z=2.4955~{\rm GeV}\,, \quad    \nonumber \\
&&    |V_{ud}| = 0.97373 \,, \quad 
    |V_{us}| = 0.2243 \, ,\quad 
    |V_{cd}| = 0.221  \, , \nonumber \\ \nonumber    
&&    |V_{cs}| = 0.975\,, \quad\, 
     |V_{cb}| = 40.8 \times 10^{-3}\,, \quad
 |V_{ub}| = 3.82 \times 10^{-3}\,, \\
&&|V_{td}| = 8.6 \times 10^{-3} \,, \quad
|V_{ts}| = 41.5 \times 10^{-3}\,, \quad
|V_{tb}| = 1.014 \,\\[0.1cm]
&&
m_u=m_d=0.33~{\rm MeV}\,, \quad
m_s=0.5~{\rm MeV}, \nonumber \\
&& m_c=1.27~{\rm GeV}\,, \quad
m_t=172.69~{\rm GeV}\,, \quad
m_b=4.18~{\rm GeV}\,, \quad \nonumber \\
&&m_e=5.11\times 10^{-4}~{\rm MeV}\,, \quad \nonumber 
m_\mu=105.66~{\rm MeV}\,, \quad
m_\tau=1.77682~{\rm GeV} \, ,\nonumber 
\eea
which are taken from the PDG~\cite{ParticleDataGroup:2022pth} except for the light quark masses which are set to be the same as in the {\sc\small Pythia8} default for consistency with its hadronisation model. 

\paragraph{Impact of NLO QCD corrections.} 
%
\begin{figure}[htbp]
    \centering
\includegraphics[width=0.44\textwidth]{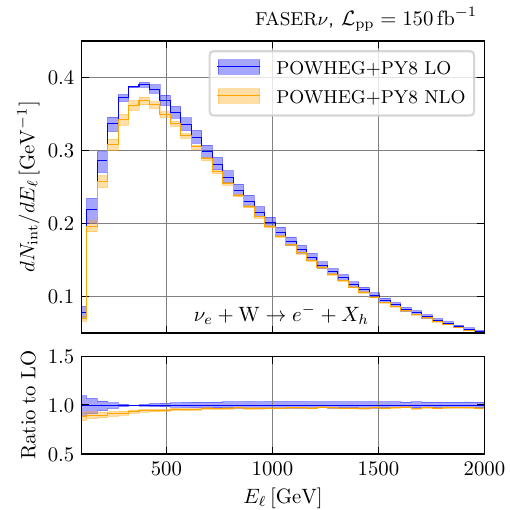}
\includegraphics[width=0.44\textwidth]{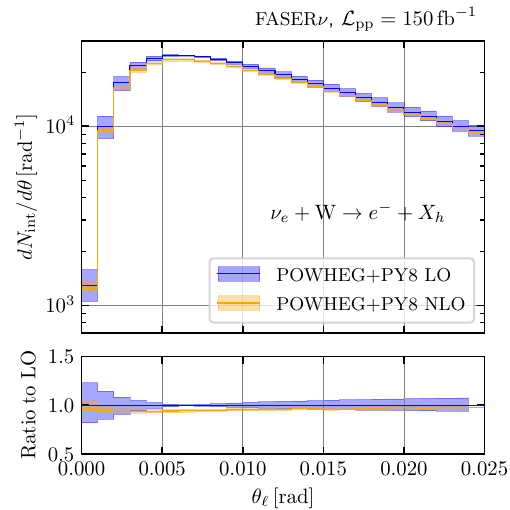}
\includegraphics[width=0.44\textwidth]{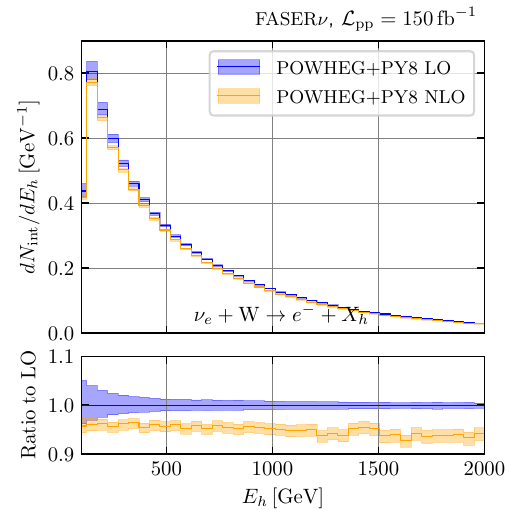}
\includegraphics[width=0.44\textwidth]{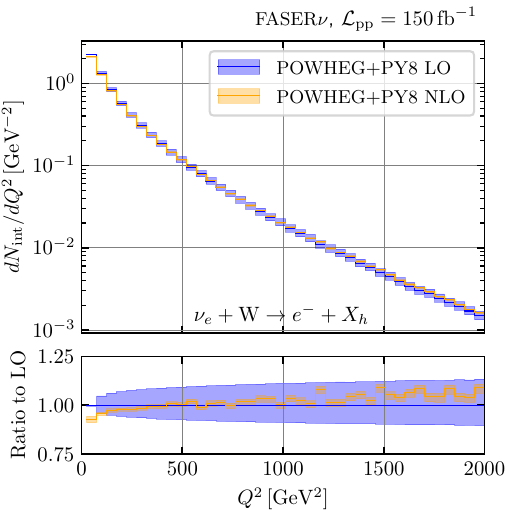}
\includegraphics[width=0.44\textwidth]{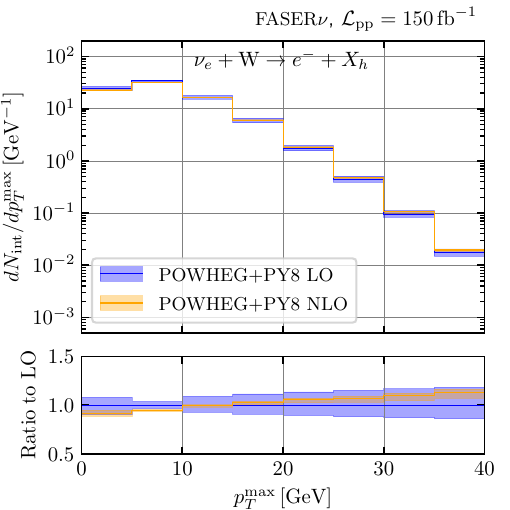}
\includegraphics[width=0.44\textwidth]{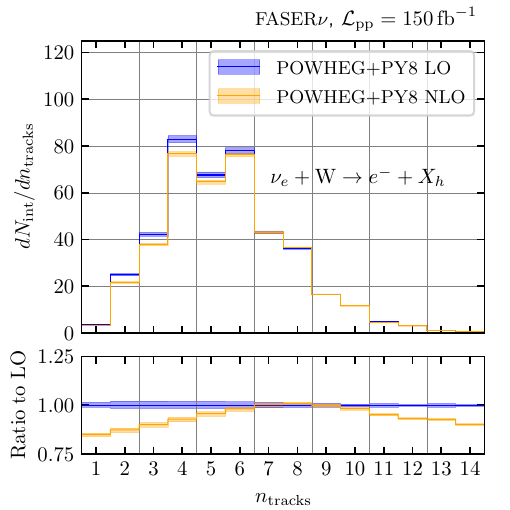}
    \caption{Differential event yields for
    representative observables in electron-neutrino CC scattering at FASER$\nu$.
    We display the charged-lepton energy $E_\ell$ and scattering angle $\theta_\ell$, the hadronic energy $E_h$, the momentum transfer $Q^2$,
    the transverse momentum of the hardest charged particle $p_T^{\rm max}$, and the charged particle multiplicity $n_{\rm tracks}$. 
    We provide comparisons of {\sc\small POWHEG}+{\sc\small Pythia8} (dipole shower) at LO and NLO accuracy together with the corresponding theory uncertainty bands from scale variations.
    The bottom panels display the ratio to the central LO prediction.
    \label{fig:faser_El_dipole}
    }
\end{figure}
%
Fig.~\ref{fig:faser_El_dipole} displays the differential event yields for the final-state charged-lepton energy, $dN^{(\nu_e)}_{\rm int}/dE_\ell$, the scattering angle, $dN^{(\nu_e)}_{\rm int}/d\theta_\ell$, the energy of the hadronic final state $dN^{(\nu_e)}_{\rm int}/dE_h$, the momentum transfer,
$dN^{(\nu_e)}_{\rm int}/dQ^2$, the  transverse momentum of the hardest particle in the event,
$dN^{(\nu_e)}_{\rm int}/dp_T^{\rm max}$, and the total charged particle multiplicity $dN^{(\nu_e)}_{\rm int}/dn_{\rm tracks}$,
in electron-neutrino CC scattering at the FASER$\nu$ detector.
We provide predictions for {\sc\small POWHEG} matched to {\sc\small Pythia8} (with the dipole shower) both at LO and NLO accuracy.
The bottom panels display the ratio of the results to the central LO prediction.
Upon integration over the NLO distributions, one recovers the total neutrino CC scattering event yields within the FASER$\nu$ acceptance listed in Table~\ref{tab:events_after_cuts}.

Here and in the rest of the section, the simulations are carried out using the NNPDF4.0 NNLO PDF set as input~\cite{NNPDF:2021njg,NNPDF:2021uiq}.
To obtain differential distributions for the corresponding nuclear target (tungsten in this case), the code is run twice, once with the proton and once with the neutron target, and the resulting distributions  are then combined bin by bin by multiplying them by the appropriate nuclear numbers $(A,Z)$ and then dividing by $A$.
Nuclear modifications of the free-nucleon PDFs are neglected, their inclusion is possible by replacing NNPDF4.0 by a nuclear PDF determination such as those from~\cite{AbdulKhalek:2022fyi,Eskola:2021nhw,Duwentaster:2022kpv}.

In Fig.~\ref{fig:faser_El_dipole} we also display the corresponding theory uncertainty bands in the LO and NLO calculations, obtained from the 7-point scale variation prescription applied to the renormalisation $\mu_R$ and factorisation $\mu_F$ scales.
We do not consider PDF uncertainties, and recall that accounting for them would be possible by means of the {\sc\small POWHEG} reweighting functionalities.
Furthermore, considering eventual variations of the incoming neutrino flux calculation would be straightforward in our framework: alternative flux calculations can be implemented as neutrino PDF replicas in the {\sc\small LHAPDF} grids and then propagated to the simulation via the same {\sc \small POWHEG} reweighting feature, in the same manner as with the regular proton or nuclear PDF uncertainties. 

From the results of Fig.~\ref{fig:faser_El_dipole} one can observe how NLO corrections modify the LO prediction by a factor between a few percent up to about 15\%, depending on the distribution. 
For instance, both for the $E_\ell$ and $E_h$ distribution, the QCD $K$-factor is negative by a few percent and mildly sensitive to the kinematics.
For the $\theta_\ell$ distributions, the $K$-factor becomes close to unity as one approaches the edge of FASER$\nu$ angular acceptance.
We also find a marked reduction of scale uncertainties from the LO to the NLO calculation, and that in general the NLO shift is contained within the scale variation error band of the LO calculation.
Two exceptions are the $E_\ell$ distribution in the low energy region and the charged track distribution $n_{\rm tracks}$.
For the latter, events with either low or high charged track multiplicity receive large NLO corrections which are not covered by the LO scale variations. 

\begin{figure}[t]
    \centering
\includegraphics[width=0.45\textwidth]{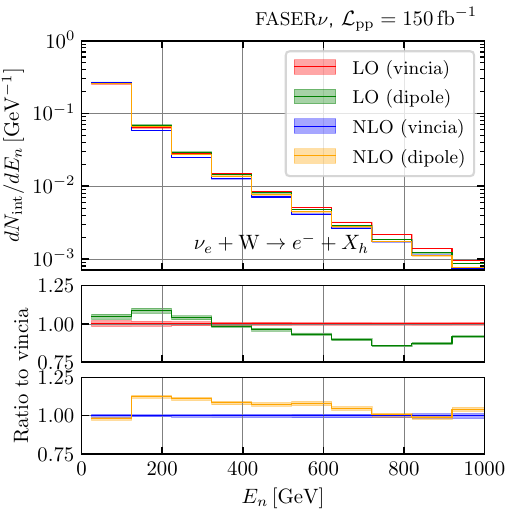}
\includegraphics[width=0.45\textwidth]{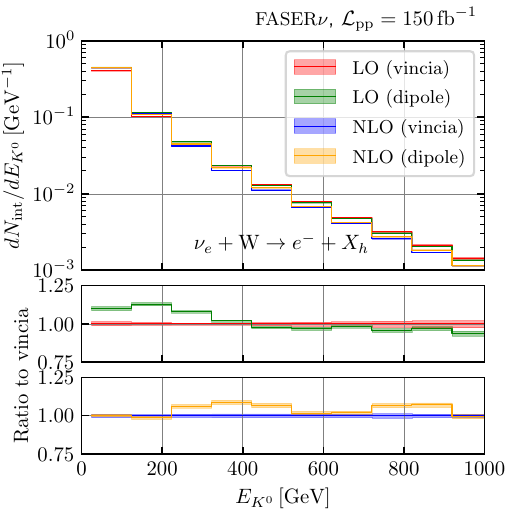}
\includegraphics[width=0.45\textwidth]{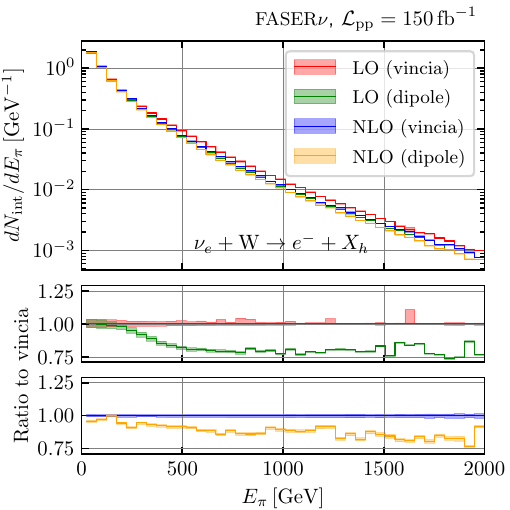}
\includegraphics[width=0.45\textwidth]{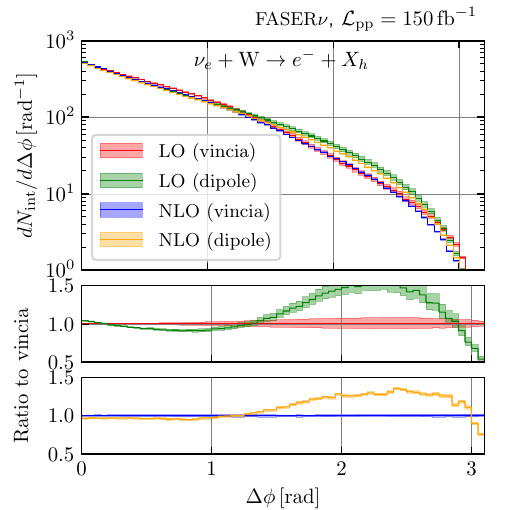}
    \caption{Same as Fig.~\ref{fig:faser_El_dipole} for the  energy of the hardest neutron ($E_n$),
    kaon ($E_K^0$), and pion ($E_{\pi}$) in the event,
    and the minimal azimuthal separation $\Delta \phi$ between the charged lepton and any other charged track in the final state. 
    We compare results obtained with two different parton showers available in {\sc\small Pythia8}, the dipole shower
    and {\sc\small Vincia}. 
\label{fig:faser_shower_dependence_inclusive}
    }
\end{figure}

\paragraph{Impact of parton shower.}
Next we assess the impact that the parton shower has on observables sensitive to the hadronic final state in the LHC far-forward neutrino experiments.
Fig.~\ref{fig:faser_shower_dependence_inclusive} shows, for electron-neutrino CC scattering at FASER$\nu$, the differential event yield distributions in the energy of the hardest neutron ($E_n$), hardest kaon ($E_{K^0}$), and hardest pion ($E_{\pi}$) in the event, as well as in the minimal azimuthal separation $\Delta \phi$ between the charged lepton and any other charged track in the final state.
Results are provided both at LO and at NLO together with the associated MHOUs.
The simulations based on the {\sc\small Pythia8} dipole shower are compared to those from the {\sc\small Vincia} shower. 
The bottom panels display the ratio between the dipole and {\sc\small Vincia}  at either LO or NLO.

In addition to the observables displayed in Fig.~\ref{fig:faser_shower_dependence_inclusive}, we have verified that distributions not sensitive to the exclusive aspects of the hadronic final state, such as $\theta_\ell$, $E_\ell$, $E_h$, or $Q^2$, do not exhibit any  significant dependence on the choice of parton shower and hence are not shown here.
Such inclusive observables are mostly independent of the pattern of QCD radiation and can be evaluated with the structure function formalism, and it is expected to that they are not affected by the modelling of the hadronic final state.

From Fig.~\ref{fig:faser_shower_dependence_inclusive} we find a marked impact of the choice of parton shower for the exclusive observables $E_\pi$, $E_n$, $E_{K^0}$ and $\Delta\phi$.
Indeed, in the {\sc\small POWHEG} NLO simulations, the choice of parton shower leads to up to a difference of up to 15\% in the high-energy tail of the $E_\pi$, $E_n$, and $E_{K^0}$ distributions, and up to 30\% difference in the $\Delta\phi$ distribution.
We note how the inclusion of NLO corrections in {\rm \small POWHEG} partially reduces the parton shower sensitivity in the $E_\pi$, $E_n$, and $E_{K^0}$ distributions, due to the fact that the hardest emission is there included with the exact matrix element.
For instance, for the $E_\pi$ distribution, NLO corrections reduce its dependence on the parton shower by up to a factor two at large $E_\pi$ values.

Furthermore, as is well known, NLO scale variations in general do not comprise the differences associated to the use of parton shower models, and hence theoretical predictions for FASER$\nu$ and other LHC far-forward neutrino experiments will display some dependence on the choice of parton shower for observables sensitive to the modelling of the hadronic final state.

\paragraph{Impact of soft QCD physics.}
In addition to the choice of parton shower model, predictions based on Monte Carlo event generators also depend on the tune adopted to describe soft and non-perturbative QCD phenomena such as hadronization, the underlying event, or multiple parton interactions. 
In~\cite{Fieg:2023kld}, {\sc \small Pythia8} was tuned to forward data from the LHCf experiment~\cite{LHCf:2015nel,LHCf:2020hjf,LHCf:2015rcj,LHCf:2017fnw} to adjust the parameters that control the modelling of the beam remnant hadronisation in the forward region.
To assess the impact of the {\sc \small Pythia8} tune in the predictions presented here, we compare this ``forward physics'' tune with the default Monash 2013~\cite{Skands:2014pea} tune used for the rest of the predictions shown in this work.
While the tune of~\cite{Fieg:2023kld} is not tailored to the neutrino DIS process, its use represents a convenient estimate of the stability of our predictions with respect to the choice of soft QCD tune in the shower.

\begin{figure}[t]
    \centering
\includegraphics[width=.44\columnwidth]{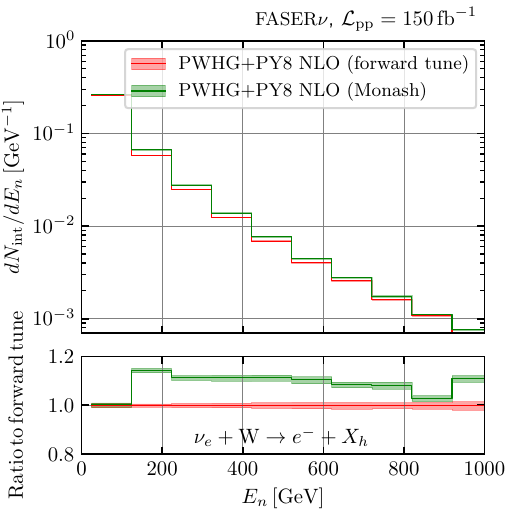}
\includegraphics[width=.44\columnwidth]{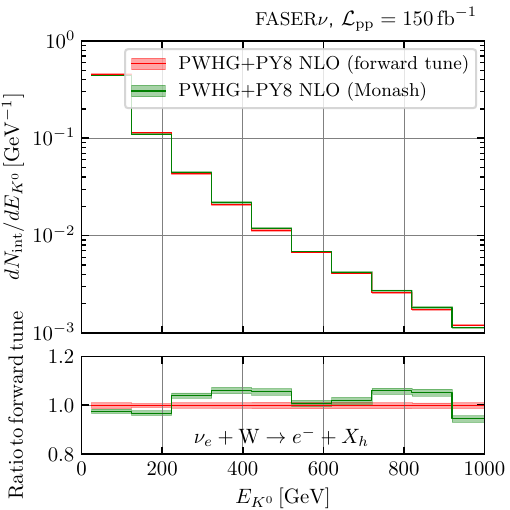}
\includegraphics[width=.44\columnwidth]{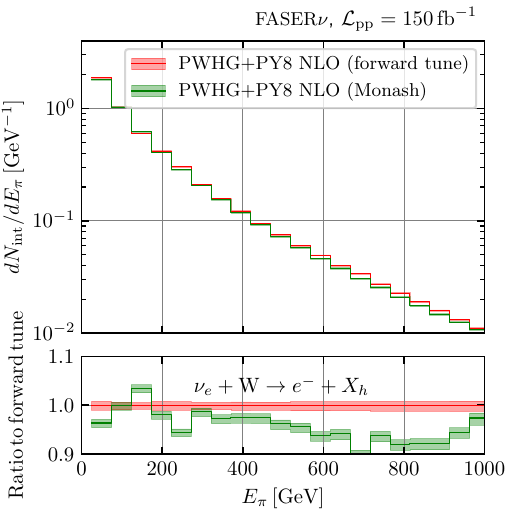}
\includegraphics[width=.44\columnwidth]{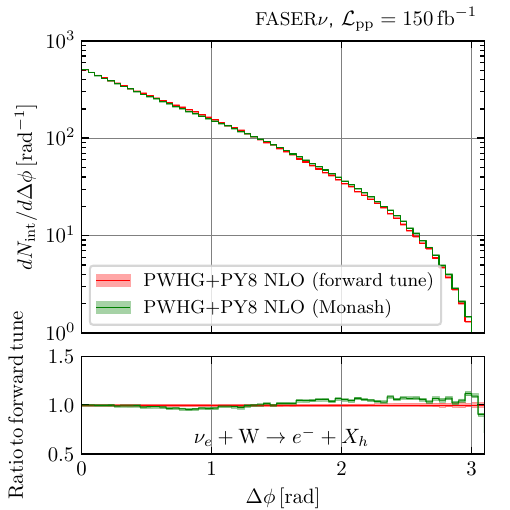}
\caption{Similar to Fig.~\ref{fig:faser_shower_dependence_inclusive}, now for the dependence of the {\sc\small POWHEG}+{\sc\small Pythia8} NLO predictions on the choice of {\sc\small Pythia8} tune, comparing Monash 2013 with the dedicated tune for forward physics from~\cite{Fieg:2023kld}.
    \label{fig:fpftune}
    }
\end{figure}

Fig.~\ref{fig:fpftune} displays the same observables of Fig.~\ref{fig:faser_shower_dependence_inclusive} now assessing the dependence of the {\sc\small POWHEG}+{\sc\small Pythia8} NLO predictions on the choice of {\sc\small Pythia8} tune, by comparing the default Monash 2013 with the tune for forward physics from~\cite{Fieg:2023kld}.
Consistently with the study of the parton shower sensitivity, inclusive observables such as $\theta_\ell$ and $E_h$ are independent of the choice of tune, while exclusive quantities such as $E_\pi$ and $\Delta\phi$ exhibit small but non-negligible differences. 
Other observables not considered here, such as charm production or hadronic flavour ratios, may be more affected by the choice of hadronisation tune.
Ultimately, a dedicated tune for soft QCD physics in neutrino DIS using the data from the LHC far-forward experiments may be developed in order to achieve the best possible description of exclusive hadronic final states.

\begin{figure}[t]
    \centering
\includegraphics[width=0.32\textwidth]{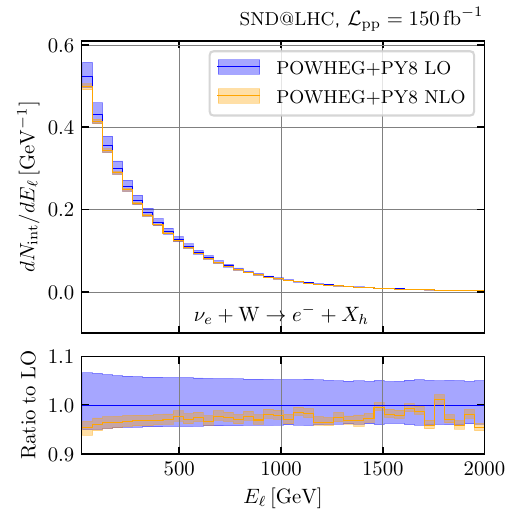}
\includegraphics[width=0.32\textwidth]{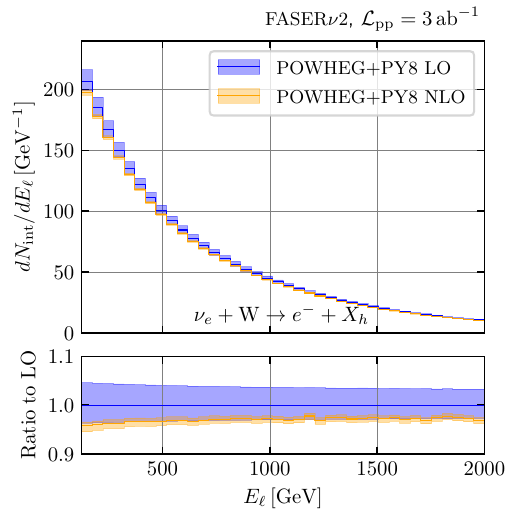}
\includegraphics[width=0.32\textwidth]{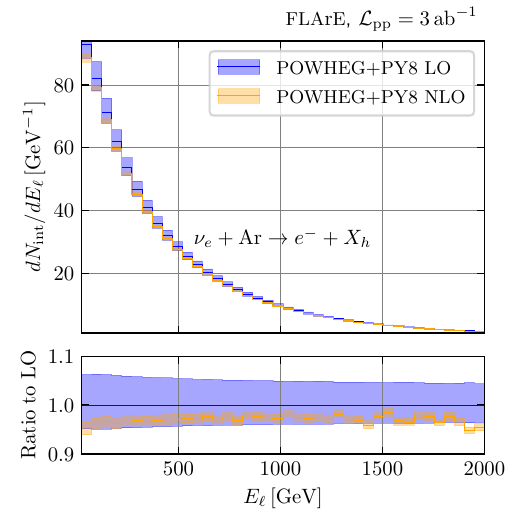}
\includegraphics[width=0.32\textwidth]{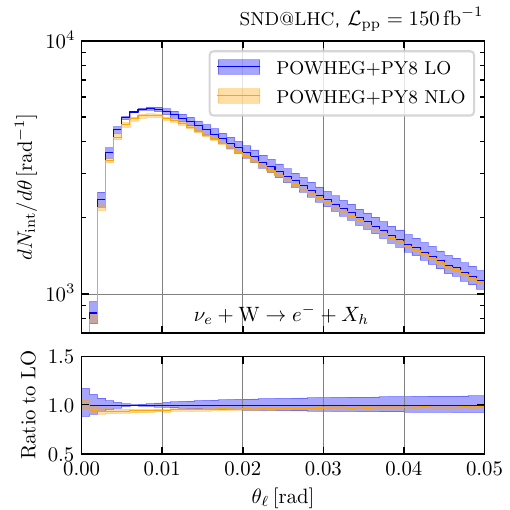}
\includegraphics[width=0.32\textwidth]{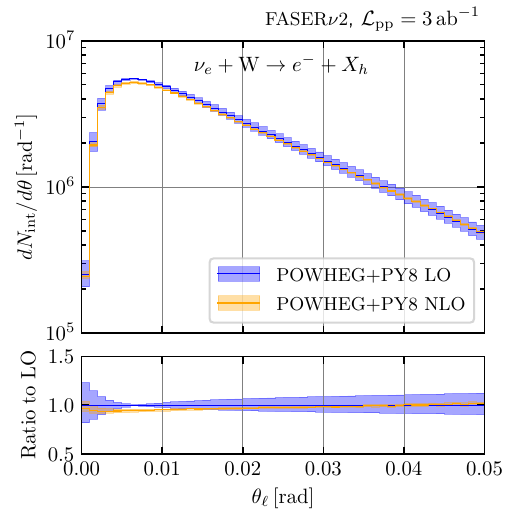}
\includegraphics[width=0.32\textwidth]{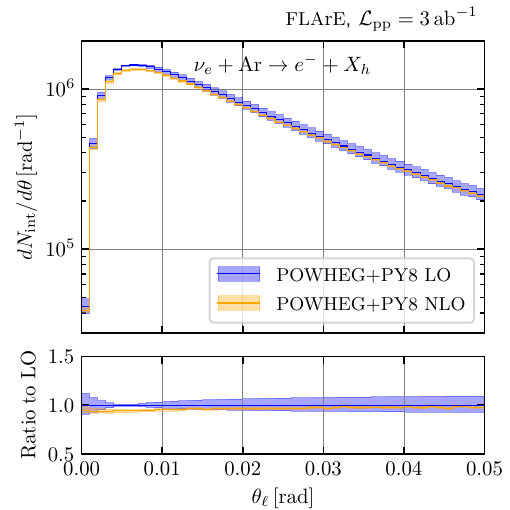}
\includegraphics[width=0.32\textwidth]{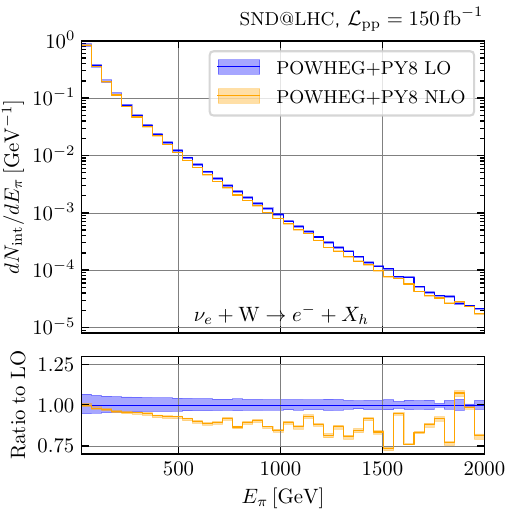}
\includegraphics[width=0.32\textwidth]{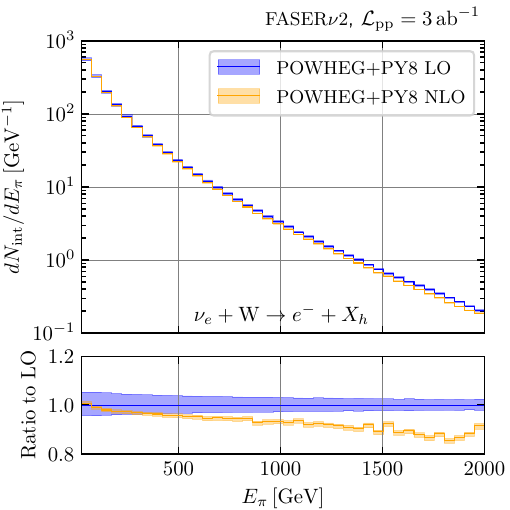}
\includegraphics[width=0.32\textwidth]{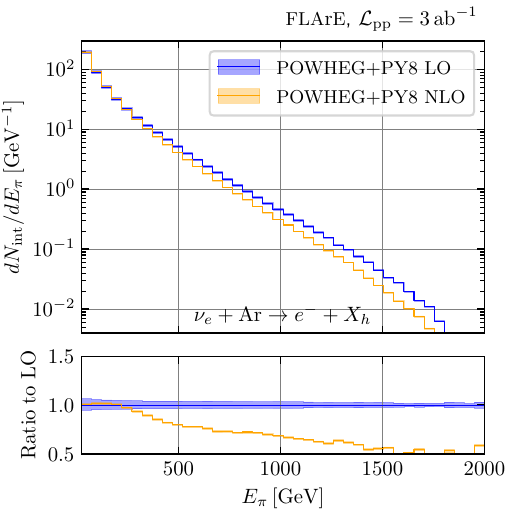}
    \caption{Same as Fig.~\ref{fig:faser_El_dipole} for the SND@LHC, FASER$\nu$2, and FLArE experiments (from left to right) for the $E_\ell$, $\theta_\ell$, and $E_\pi$ differential distributions (from top to bottom).
    \label{fig:exp_dependence}
    }
\end{figure}

\paragraph{Results for SND@LHC and the FPFs.}
The previous discussion focused on the {\sc\small POWHEG+Pythia8} predictions for FASER$\nu$.
Now we present representative predictions for SND@LHC as well as for two of the proposed FPF experiments, namely FASER$\nu$2 and FLArE.
Fig.~\ref{fig:exp_dependence} displays a similar comparison as in Fig.~\ref{fig:faser_El_dipole}, now for SND@LHC, FASER$\nu$, and FLArE,
considering the $E_\ell$,  $\theta_\ell$, and $E_\pi$ distributions.
Comparing the results for these three experiments with corresponding predictions for FASER$\nu$, one observes that the qualitative behaviour of both the observables themselves and of the NLO QCD corrections are qualitatively similar in all cases, with the major difference being the overall normalisation (see Table~\ref{tab:events_after_cuts}).
This finding that the qualitative impact of NLO QCD corrections, the role of parton shower, and the effects of the {\sc\small Pythia8} tune are to first approximation independent of the specific LHC far-forward experiment considered can be partially understood by the similar shapes of the neutrino PDFs displayed in Fig.~\ref{fig:neutrino-pdfs-exps}.

\begin{figure}[t]
    \centering
\includegraphics[width=.46\columnwidth]{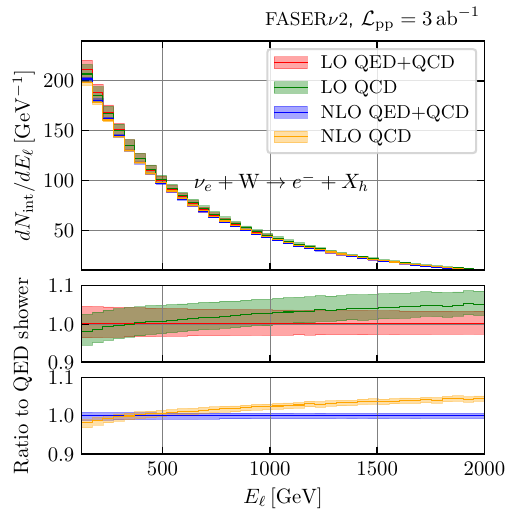}    \includegraphics[width=.46\columnwidth]{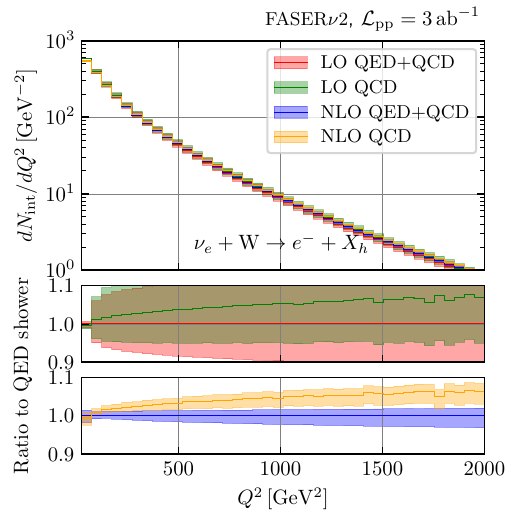}
\includegraphics[width=.46\columnwidth]{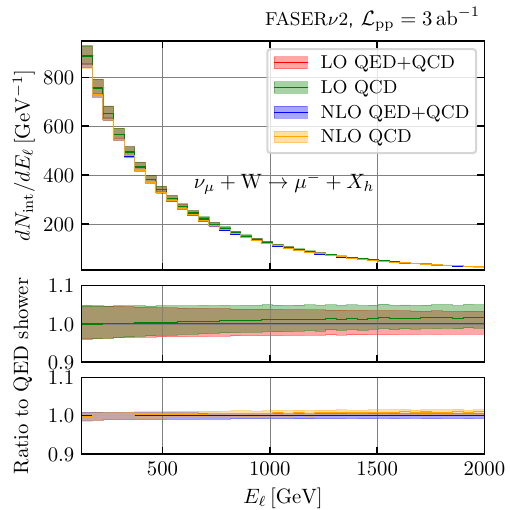}    \includegraphics[width=.46\columnwidth]{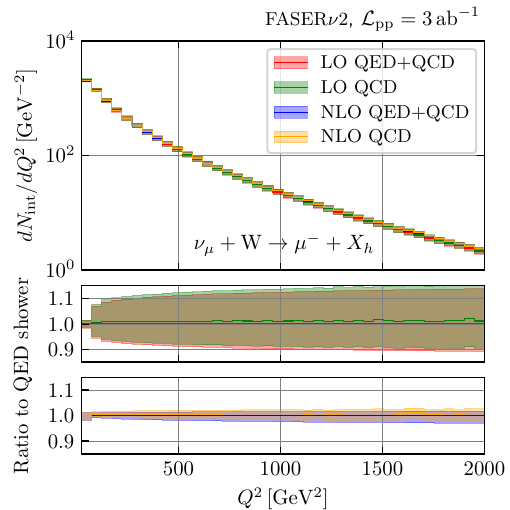}
    \caption{Differential distributions in $E_\ell$ and $Q^2$ for bare electrons (top) and muons (bottom panels) produced in CC scattering at FASER$\nu$2. 
    The curves labelled ``QCD'' were obtained using the dipole shower in {\sc \small Pythia8}, while for the ``QED+QCD'' curves also the QED shower was included in the simulation (NLO virtual QED corrections to the matrix element are not considered).
    }
    \label{fig:qedonoff}
\end{figure}

\paragraph{QED shower effects.} 
The matching of {\sc\small POWHEG} to {\sc\small Pythia8} makes it possible to account for the effects of QED showers in the Monte Carlo simulations of the LHC far-forward neutrino experiments.
To illustrate these capabilities, Fig.~\ref{fig:qedonoff} displays for FASER$\nu$2 the differential distributions in $E_\ell$ and $Q^2$ in {\sc\small POWHEG+Pythia8} without and with the QED shower included, for the case of bare charged leptons. 
NLO virtual QED corrections to the matrix elements are not considered here.
QED shower effects are more important for bare electrons, specially at low-$E_\ell$ and high-$Q^2$ where they reach the 10\% level, while they are smaller for bare muons. 
We note that QED radiation off the final-state charged lepton may modify whether some events satisfy the acceptance requirements in $E_\ell$ and $\theta_\ell$.
We have verified that observables sensitive to the details of the hadronic final state, such as $E_\pi$ and $\Delta \phi$ ,remain unaffected by the inclusion of QED radiation in the shower.

The observed dominant effects of QED radiation for bare leptonic observables can be captured by dressing the charged leptons with photons and leptons lying in a cone of $\Delta R=\sqrt{\Delta \eta ^2 +\Delta \phi ^2} =0.2$ around the hardest lepton of the event.
In the case of such dressed observables, we checked that the QED corrections shown  in Fig.~\ref{fig:qedonoff} are reduced down to the 1\% level at most. 

\paragraph{Dependence on the neutrino flavour.}
\begin{figure}[htbp]
    \centering
\includegraphics[width=0.46\textwidth]{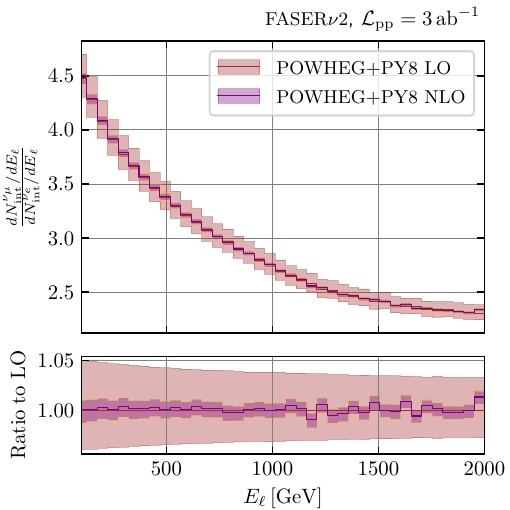}
\includegraphics[width=0.46\textwidth]
{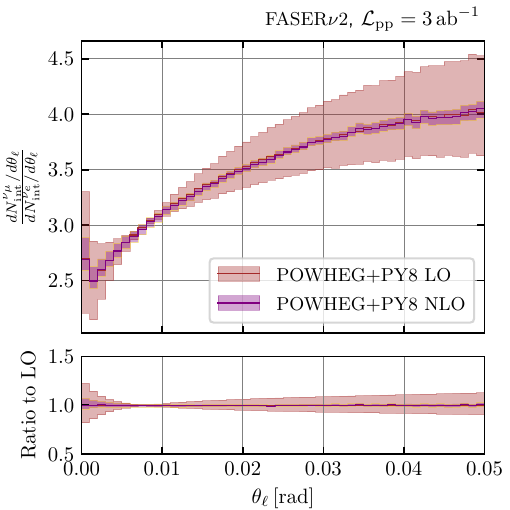}
\includegraphics[width=0.46\textwidth]{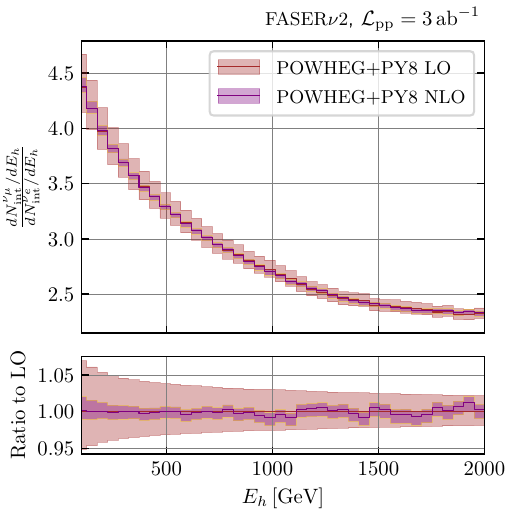}
\includegraphics[width=0.46\textwidth]{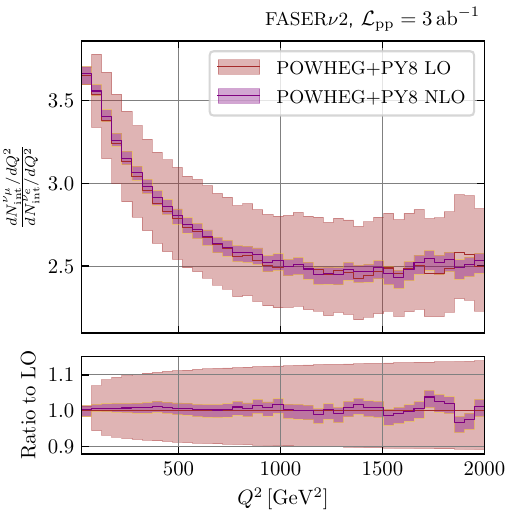}
\includegraphics[width=0.46\textwidth]{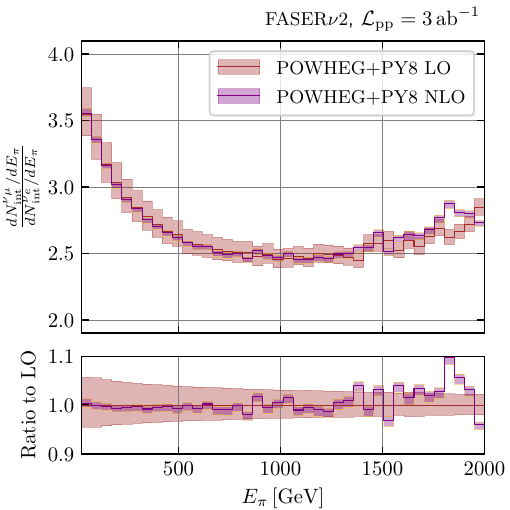}
\includegraphics[width=0.46\textwidth]{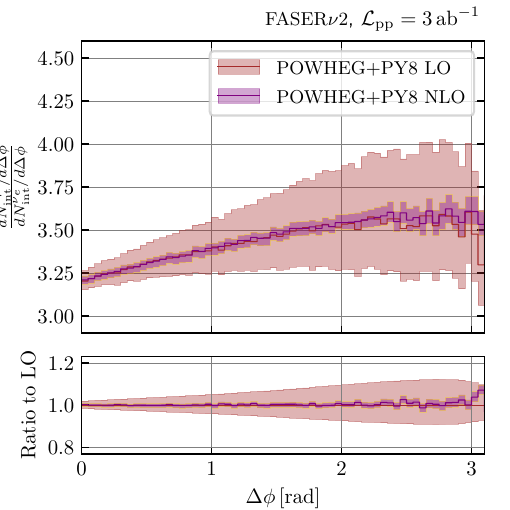}
    \caption{The ratio of event yields evaluated between muon-neutrino and electron-neutrino CC scattering for the $E_\ell$, $\theta_\ell$, $E_h$, $Q^2$, $E_\pi$, and $\Delta\phi$ distributions in the case of the FASER$\nu$2 detector.
    We provide {\sc\small POWHEG+Pythia8} predictions both at LO and NLO together with the corresponding theory uncertainty bands (MHOUs) in each case.
    }
    \label{fig:faserv2numuvsnue}
\end{figure}
%
We now assess the dependence of the {\sc\small POWHEG} NLO predictions with respect to the flavour of the incoming neutrino beam.
To isolate this dependence, we evaluate ratios of event yields between muon-neutrino and electron-neutrino CC scattering events, such as for the charged-lepton energy
\be
\label{eq:neutrino_flavour_ratios}
R_{\nu_\mu/\nu_e}(E_\ell)\equiv \frac{d N_{\rm int}^{(\nu_\mu)}(E_\ell)}{dE_\ell}\Bigg/ \frac{d N_{\rm int}^{(\nu_e)}(E_\ell)}{dE_\ell} \, ,
\ee
and likewise for other final-state observables and different neutrino flavour combinations.

The neutrino flavour ratios defined in Eq.~(\ref{eq:neutrino_flavour_ratios}) are particularly attractive for many phenomenological applications. 
For instance, they provide information on possible violations of lepton-flavour universality in the neutrino sector.
Furthermore, theoretical uncertainties in the production mechanisms partially cancel out.
This feature is best illustrated by the ratio $R_{\nu_\tau/\nu_e}(E_\nu)$ at high neutrino energies, for which QCD uncertainties associated to the dominant $D$-meson production and fragmentation mechanisms partially cancel out while retaining sufficient sensitivity to the small-$x$ gluon PDF.

Fig.~\ref{fig:faserv2numuvsnue} displays the ratio of the event yields evaluated between muon-neutrino and electron-neutrino CC scattering events, Eq.~(\ref{eq:neutrino_flavour_ratios}),  for the $E_\ell$, $\theta_\ell$, $E_h$, $Q^2$, $E_\pi$, and $\Delta\phi$ distributions in the case of the FASER$\nu$2 experiment.
We provide the {\sc\small POWHEG+Pythia8} predictions both at LO and NLO, together with the corresponding theory uncertainty bands (from MHOUs) in each case. 
The ratios $R_{\nu_\mu/\nu_e}$ are found to range between a factor 2 and 8, depending on the observable and the kinematics, reflecting the higher muon-neutrino fluxes in comparison with the electron-neutrino ones and consistently with the behaviour of neutrino PDFs shown in Fig.~\ref{fig:neutrino-pdfs-flavour}.

From the neutrino flavour ratio observables of Fig.~\ref{fig:faserv2numuvsnue}, one also observes that the central prediction is stable with respect to the addition of NLO QCD corrections.
This feature can be understood from the fact that QCD effects are flavour blind.
Hence the magnitude and shape of these ratio observables mostly reflect the differences at the level of incoming fluxes and are not sensitive to higher-order QCD corrections.
One also notes how for these observables the theory uncertainties from scale variations are markedly reduced when going from LO to NLO accuracy. 
The availability of precise predictions for these ratio observables motivates their deployment in experimental analyses, where many systematics will also cancel out, to study the mechanisms of forward hadron production at the LHC and to search for deviations from the SM interactions in the neutrino sector at TeV energies.

The results of Fig.~\ref{fig:faserv2numuvsnue} also provide relevant information that may inform experimental analyses, in particular concerning the event selection criteria. 
For instance, the minimal azimuthal angle $\Delta \phi$ between the final state lepton and the closest charged track in the system indicates that muons from CC interactions will be characterised by a larger typical separation in azimuthal angle as compared to electrons.
Along similar lines, one can select muon-neutrino enriched samples by applying cuts in the $E_h$ or $Q^2$ distributions, to restrict the analysis regions to the kinematics where  $R_{\nu_\mu/\nu_e}$ is the largest.
While such considerations are not necessary for detectors such as FASER$\nu$ which can identify the neutrino flavour on an event-by-event basis, they may be of assistance for FASER during Run-4 where only the electronic components of the detector will be available.

\section{Summary and outlook}
\label{sec:summary}

In this work, we have presented a detailed phenomenological study of final-state kinematic distributions relevant for the interpretation of present and future LHC far-forward neutrino experiments.
The analysis is carried out  at NLO accuracy in the QCD expansion and matched to a modern parton shower generator such as {\sc\small Pythia8}.
Within our approach, based on the {\sc\small POWHEG-BOX-RES} event generator, the incoming neutrino fluxes are parameterised in terms of a neutrino PDFs making the description of $\nu p$ scattering analogous to that of $pp$ collisions.
As demonstrated here, the inclusion of higher-order QCD corrections in neutrino Monte Carlo simulations is necessary for a reliable estimate of the associated theoretical uncertainties in these predictions.

We have quantified the impact of NLO corrections to differential observables within the acceptances of the FASER$\nu$ and SND@LHC (Run-3), and 
FASER$\nu$2 and FLArE (FPF operating at the HL-LHC) experiments, showing that these effects are in general sizable and cannot be neglected.
We have also studied the impact of varying the parton shower model and its tune for hadronisation, finding that differences for exclusive observables such as the leading pion kinematics are potentially large and not covered by the NLO scale variations.
The neutrino PDF formalism presented in this work is suitable for integration in other event generators, such as {\sc\small Herwig7}~\cite{Bellm:2015jjp,Bewick:2023tfi} and {\sc\small Sherpa}~\cite{Sherpa:2019gpd}, which provide stand-alone predictions for DIS processes at NLO+PS accuracy.

The availability of {\sc\small POWHEG}-based generators tailored to the simulation of LHC neutrinos~\cite{FerrarioRavasio:2024kem,Buonocore:2024pdv} opens the door to many possible novel experimental analyses relying on precise theory simulations.
To begin with, FASER will operate during Run-4 without the emulsion component (FASER$\nu$).
Being able to continue its successful neutrino programme hence demands a robust modelling of the leptonic and hadronic final states in neutrino scattering and their interactions with the different components of the electronic detector and the calorimeters, which can be provided by the {\sc\small POWHEG} simulations.
Along similar lines, this tool may be used to assess the feasibility of measuring neutral-current neutrino scattering (where only the hadronic final state is reconstructed), which in turn would enable a direct measurement of the Weinberg angle $\sin^2 \theta_W$ at TeV energies.
As a third possible application, {\sc\small POWHEG} simulations may be used to relate experimental observables (neutrino event yields) to the incoming neutrino fluxes to attempt a direct determination of the neutrino PDFs from the far-forward detectors data.
Such a measurement of the neutrino PDFs would result in stringent constraints on the mechanisms governing forward hadron production at the LHC and provide key inputs for astroparticle physics, such as the first laboratory-based determination of the prompt flux for neutrino telescopes~\cite{Gauld:2015yia}.

As already highlighted in~\cite{FerrarioRavasio:2024kem} for the fixed neutrino energy case, beyond applications to the LHC the {\sc\small POWHEG} framework can also be applied to the modelling of DIS processes at neutrino observatories and related experiments.
Indeed, provided the corresponding atmospheric and astrophysical neutrino fluxes are provided in the {\sc\small LHAPDF} neutrino PDF format, our simulations can be extended to model CC and NC deep-inelastic scattering at
oscillation and astroparticle physics experiments such as DeepCore~\cite{IceCube:2017lak} or KM3NET/ORCA~\cite{KM3NeT:2021ozk} and IceCube~\cite{IceCube:2014stg} or KM3NET/ARCA~\cite{KM3Net:2016zxf}.
Robust QCD predictions for these experiments are specially relevant for analyses sensitive to the details of the hadronic final state, such as those based on neutral-current scattering signatures. 

The results of this work represent an important contribution to the exciting program of precision QCD and neutrino physics at the LHC far-forward detectors, by providing a flexible simulation framework based on higher-order QCD corrections, precise PDFs, and modern parton shower and hadronisation algorithms.

\subsection*{Acknowledgements}

We are indebted to Rhorry Gauld, Barbara Jager, Alexander Karlberg,  and Giulia Zanderighi for productive interactions concerning the {\sc\small POWHEG-BOX-RES} DIS code.
We are grateful to Alfonso Garcia for discussions on the topic and for assistance with the predictions from
the {\sc\small GENIE} event generator.
We thank Felix Kling and Luca Rottoli for discussions concerning the LHC neutrino fluxes, and Anna Sfyrla for suggestions concerning relevant observables at FASER. 

The work of J.~R. is partially supported by NWO, the Dutch Research Council, and by the Netherlands eScience Center (NLeSC).
E.~G. acknowledges the financial support for the Olga Igonkina Scholarship, and is grateful to Marco Bonvini for hospitality in Rome while this work was being completed.

\bibliographystyle{utphys}
\bibliography{nuPOWHEG}

\end{document}